\documentclass[preprint,12pt]{elsarticle}
\usepackage{amsmath,amssymb,amsfonts}
\usepackage{graphicx}
\usepackage{subcaption}
\usepackage{booktabs}
\usepackage{siunitx}
\usepackage{pdflscape}
\usepackage{makecell}
\usepackage{array}
\usepackage{tabularx}
\usepackage[table,xcdraw]{xcolor}
\usepackage{cleveref}
\usepackage{float}
\usepackage{textcomp}
\usepackage{stfloats}
\usepackage{url}
\usepackage{verbatim}
\usepackage{natbib}
\usepackage[a4paper, margin=1in]{geometry}
\usepackage{cleveref}
\journal{Reliability Engineering \& System Safety}

\begin{document}

\begin{frontmatter}

\title{Managing the unexpected: Operator behavioural data and its value in predicting correct alarm responses}

\author[1,2]{Chidera W. Amazu}
\author[2,3]{Joseph Mietkiewicz}
\author[1,2]{Ammar N. Abbas}
\author[1]{Gabriele Baldissone}
\author[1]{Davide Fissore}
\author[1]{Micaela Demichela}
\author[3,4]{Anders L. Madsen}
\author[2]{Maria Chiara Leva}

\address[1]{Politecnico di Torino, Turin, Italy}
\address[2]{Technological University Dublin, Dublin, Ireland}
\address[3]{Hugin Expert A/S, Aalborg, Denmark}
\address[4]{Aalborg University, Aalborg, Denmark}

\begin{abstract}
Data from psychophysiological measures can offer new insight into control room operators' behaviour, cognition, and mental workload status. This can be particularly helpful when combined with appraisal of capacity to respond to possible critical plant conditions (i.e. critical alarms response scenarios). However, wearable physiological measurement tools such as eye tracking and EEG caps can be perceived as intrusive and not suitable for usage in daily operations. Therefore, this article examines the potential of using real-time data from process and operator-system interactions during abnormal scenarios that can be recorded and retrieved from the distributed control system's historian or process log, and their capacity to provide insight into operator behavior and predict their response outcomes, without intruding on daily tasks. Data for this study were obtained from a design of experiment using a formaldehyde production plant simulator and four human-in-the-loop experimental support configurations. A comparison between the different configurations in terms of both behaviour and performance is presented in this paper. A step-wise logistic regression and a Bayesian network models were used to achieve this objective. The results identified some predictive metrics and the paper discuss their value as precursor or predictor of overall system performance in alarm response scenarios. Knowledge of relevant and predictive behavioural metrics accessible in real time can better equip decision-makers to predict outcomes and provide timely support measures for operators.
\end{abstract}

\begin{keyword}
Operational data \sep Bayesian Network \sep Logistic Regression \sep Decision support \sep Control rooms \sep Alarm handling \sep Human factorssiri
\end{keyword}

\end{frontmatter}

\section{Introduction}
Operators in modern process control room configurations play a crucial role in preventing accidents and managing unexpected events in contemporary process plants. Typically, operators in these settings monitor plant processes, manage alarms, diagnose issues, and execute appropriate control actions. However, they face several challenges that hinder their ability to perform these roles effectively. One key challenge arises from the tools designed to support their tasks. More specifically, tools such as alarm systems, procedures and displays, which are intended to enhance their situational awareness, decision making, and ability to manage future incidents, may not always provide adequate support \cite{Simonson2022},\cite{Gao2013}. 

In addition, in more advanced automated plants, operators face a new challenge: extended periods of monitoring, during which their roles may be partially or entirely assumed by automated systems, leaving them out of the loop \cite{Hinss2022}. The problem with this shift is that, when these automated systems fail, operators who have been disengaged suddenly become critical when an alarm is triggered. This lack of situational awareness or the inability to make correct decisions due to insufficient support can be catastrophic in safety-critical environments. These issues underscore the necessity of adopting a predictive approach to anticipate factors that could impede correct alarm responses by operators and to develop strategies to manage such situations. Predictive insights can be achieved using machine learning or deep learning techniques. However, a fundamental step is to identify the variables required for these predictions, which accurately describe the operator-system interaction.

Several performance shaping factors have been recognised as influential in evaluating the operator-system interaction in process control, most of which are based on traditional human reliability analysis methods, such as SPAR-H \cite{Gertman2004}. Some researchers have also identified psycho-physiological measures that describe these interactions, exploring their use in predicting operators’ workload levels \cite{Iqbal2024}, situational awareness, or comparing different support options. However, these approaches have limitations. Human reliability analysis methods are typically calculated either before or after initiating events, require multiple steps and expert assessment, and do not provide objective, context-specific data. On the other hand, psycho-physiological measures, such as those obtained by eye tracking, pose problems such as intrusiveness of the device, lack of applicability in real-time, and the need for controlled setups \cite{Braarud2021}. Therefore, the goal is to identify real-time measures that describe the operator-system interaction and can help predict an operator’s ability to respond correctly to alarms. This would enable the provision of timely, real-time recommendations to operators, enhancing safety and efficiency in process control environments. 

Operational data, specifically those that describe the human system interaction, such as information on the acknowledgement of alarms, response times, number of alarms, etc., are useful to decision makers as they have the potential to provide insights into metrics for understanding operator behaviour, workload, situational awareness, and errors \cite{Rahman2020}, \cite{Braarud2021}, \cite{Chen2012}. They provide practical and unbiased information. The research of Braarud and Chen specifically highlighted the value of real-time operator-system interaction data for classifying the operator workload \cite{Braarud2021}, \cite{Chen2012}. Their studies identified important features such as the number of alarms per station, human-system interaction (HSI) activities such as navigating process images, selecting objects, and entering values, along with speech characteristics. 

In process plants, obtaining the necessary operational data presents several challenges. For example, recovery from safety or load-related alarms is typically not automated due to the wide variety of potential causes. These manual interventions are logged in audit trails and stored in historians or databases, yet they are rarely analysed, resulting in a blind spot in digitisation and Industry 4.0 initiatives \cite{9557368}. To address this, human-system interaction data from historians should be examined; however, little research has explored these data for error prediction or as potential indicators of process safety and control performance \cite{CROMPTON202183}.

This gap may stem from the limitations in how historian data are recorded and their original intended purpose. Even in cases where such data are available, as demonstrated in a previous analysis of a gas exploration and production (E \& P) company, the methods used to record the data can pose significant challenges during analysis, making it difficult to extract meaningful behavioural metrics.

\subsection{Research Questions}
Against this background, this article aims to address  the following questions:
\begin{enumerate} 
    \item  Is it possible to gain useful insight on the dynamic of human machine interaction during alarm response conditions that can be used to predict the probability of a successful resolution for the alarmed state?
    \item What operator behavioural metrics are the best predictor for the successful recovery of a plant-alarmed state? 
    \item Which of the support configurations tested in this study is more likely to have a positive impact on the operator's ability to successfully manage plant-alarmed conditions?
\end{enumerate}

\subsection{Brief outline of the paper}

This paper will present briefly the performance and behavioural outcomes from an experimental study that compare four human-in-the-loop configurations using the operational data derived by running a Design of experiment involving 92 part icipants managing three alarm response scenarios using the HMI developed for the simulator of a formaldehyde production plant (REF). 

Then, the essential behavioural measures from the fours configurations are selected for further prediction of success and failure outcomes. A two-step approach to predict control room operator errors, leveraging both Logistic Regression and Bayesian Networks is employed to create robust and insightful predictive models. Using these two techniques allows one to harness the strengths of each technique, providing a comprehensive framework for understanding and mitigating errors in safety-critical environments.

Initially, Logistic Regression (LR) is used to identify and test the significance of various predictor variables. Logistic Regression is a powerful statistical method well-suited for binary classification tasks. Provides clear and interpretable results that show the impact of each predictor variable on the outcome. The simplicity and efficiency of LR make it an ideal starting point for the analysis, especially when dealing with large datasets. Using stepwise logistic regression, we iteratively select the most significant variables, ensuring that the final model is both streamlined and robust. This step is crucial because it helps to understand the relative importance of different predictors and provides a solid foundation for further analysis. The method is then used to predict error with inputs from the behavioural data and subsequently with subjective measures.

Following the initial analysis with Logistic Regression, Bayesian Networks (BNs) are used to model the complex dependencies among the identified significant variables. Bayesian Networks are particularly effective in capturing and quantifying the conditional dependencies between variables, making them ideal for understanding the intricate interactions within operational data. Moreover, BNs offer a significant advantage in handling missing data, a common challenge in real-world applications. Their probabilistic framework allows for inference even with incomplete datasets, enabling predictions based on available information.  This flexibility makes BNs well-suited for online deployment in control rooms, where real-time assessment of error likelihood is crucial. The model can adapt to varying levels of data completeness, providing continuous risk assessment and potentially enhancing operator performance and system safety.

\section{Material and Methods}
\subsection{Case Study}

The case study examines the production of formaldehyde through the partial oxidation of methanol with air. A secondary reaction, which fully oxidises methanol to carbon monoxide, reduces the formaldehyde yield. This study, originally presented by \cite{Demichela2017} and later adapted by the authors, details a plant consisting of six main sections.

\begin{figure}[ht]
    \centering
    \includegraphics[width=0.9\textwidth]{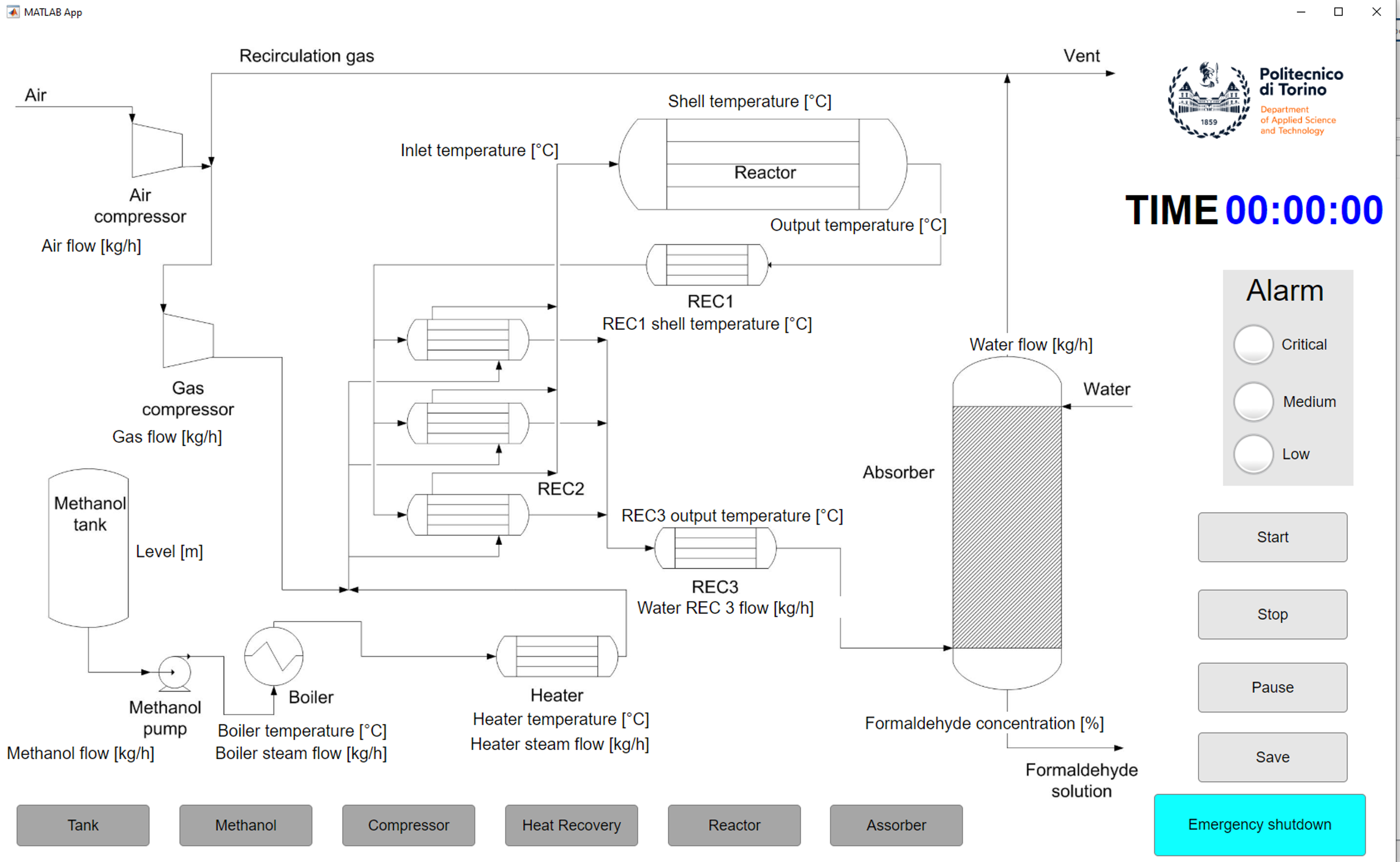}
    \caption{Main screen \cite{AmazuDOE2024}.}
    \label{fig:sim_main}
\end{figure}

As shown in \Cref{fig:sim_main}, the plant layout begins with the Tank Section, where a methanol storage tank, equipped with alarms for parameter deviations, is located. The Methanol Section follows, where liquid methanol is converted to gas via a pump and heater before mixing with compressed gas from the Compressor Section. This mixture then passes through a heat exchanger (REC2), heating it to approximately 200°C before entering the reactor. The Reactor Section contains the reactor, and the heat exchanger section includes the heat exchangers REC1, REC2, and REC3. REC1 and REC3 cool the reactor's output, and the heat recovered from REC1 and REC3 is used for water boiling and cooling, respectively.

The reactor, located in the Reactor Section, works in conjunction with the heat exchangers in the Heat Exchanger Section. REC1 and REC3 cool the reactor's product, with the recovered heat used to boil and cool water. REC3 specifically cools the product to around 67°C before it enters the absorber, where it is absorbed by water flowing countercurrently. This absorber is located in the Absorber Section. Navigation through these plant sections is facilitated by buttons at the bottom of the interface display (refer to \Cref{fig:sim_main}).

Several potential accident scenarios could occur in this plant, such as overheating of the reactor or absorber if the temperatures from the mixture or product exceed expectations. Attempting to control this temperature by increasing the water input into the absorber could result in a loss of the volume of the target product.

\subsection{Participants}
The study was designed to capture conditions and process events similar to those in a formaldehyde production plant as closely as possible. Ninety-two participants voluntarily participated in this study \cite{AmazuDOE2024}. They consisted of junior process engineers selected from a pool of chemical engineering master students and some staff familiar with the plant process. Their mean age was 25 years (SD = 5.4), ranging from 21 to 61 years, with training and experience in Chemical Engineering and, in some cases, knowledge of control room operations.    Participants rated their level of expertise in both chemical process and control room operations prior to the test. They were also split equally into four groups (23 participants per group) to assess different levels of support in a simulated control room environment. This division allowed us to systematically assess the impact of various support features on operator performance, workload, and decision-making processes. Each group had a unique configuration to explore the effectiveness of different technological tools. The characteristics of each group are summarised in Table \ref{tb:group}.
 The study was reviewed and approved by the Technological University Ethics Review Committee and the CISC internal ethics committee and was performed with appropriate consent obtained from participants.
\begin{enumerate}
    \item \textbf{Group 1 - Baseline without Alarm Rationalization}: This group operated under standard conditions without any alarm rationalization system, serving as the control group.
    
    \item \textbf{Group 2 - Introduction of Alarm Rationalization}: Equipped with an alarm rationalization system, this group aimed to assess the impact of filtering out non-critical alarms on reducing cognitive load and improving focus on critical issues.
    
    \item \textbf{Group 3 - Transition to On-Screen Procedures}: This group shifted from paper-based methods to an on-screen interface for accessing procedures. The objective was to determine if digital access could enhance operational efficiency and response times compared to traditional methods.
    
    \item \textbf{Group 4 - Integration of AI DSS}: Representing the most advanced level of support, this group incorporated an AI decision support system to measure its incremental benefits in improving performance, reducing errors, and enhancing overall safety and efficiency \cite{mietkiewicz2024enhancing}.
\end{enumerate}

\begin{table}[ht]
\centering
\caption{Characteristics of the Groups/type of support}
\label{tb:group}
\begin{tabular}{|p{0.5cm}|p{3cm}|p{2cm}|p{1.8cm}|}
\hline
 & \textbf{Alarm \newline Rationalization} & \textbf{Procedure \newline On Screen} & \textbf{AI \newline Support} \\ \hline
\textbf{G1} & no & no & no \\ \hline
\textbf{G2} & yes & no & no \\ \hline
\textbf{G3} & yes & yes & no \\ \hline
\textbf{G4} & yes & yes & yes \\ \hline
\end{tabular}
\end{table}


\subsubsection{Scenarios/Task Complexity}

We designed three scenarios to evaluate different human-in-the-loop (HITL) configurations in a simulated control room environment. These scenarios aimed to test the robustness of the system and the operator's ability to respond effectively under varying conditions. Each scenario simulates a specific failure or challenges that an operator might encounter in a plant environment.

\begin{enumerate}
\item \textbf{Pressure Indicator Control Failure (Scenario 1)}: The automatic pressure management system in the tank fails, requiring the operator to manually adjust the nitrogen inflow to maintain pressure. The scenario becomes more challenging when the nitrogen flow stops, leading to a pressure drop as methanol continues to be pumped into the plant.

\item \textbf{Nitrogen Valve Primary Source Failure (Scenario 2)}: The primary nitrogen supply system fails, and the operator must switch to a backup system. The backup system starts slowly, necessitating careful regulation of pump power to maintain tank pressure.

\item \textbf{Temperature Indicator Control Failure in the Heat Recovery Section (Scenario 3)}: There's a risk of the reactor overheating, causing a pressure increase. The operator's goal is to prevent the activation of the pressure switch (PSL01) within 18 minutes following the initial alarm by managing the reactor’s temperature. This scenario corresponds to a more challenging situation with alarm overflow.
\end{enumerate}

Each scenario was designed to evaluate the effectiveness of the decision support system in aiding operators during complex and high-pressure situations. These scenarios as summarized in \Cref{tab:scenarios}, provide a comprehensive test of the system's utility in enhancing operator performance and decision-making under various challenging conditions. 

\begin{table}[ht]
\centering
\caption{Scenario Characteristics}
\label{tab:scenarios}
\begin{tabular}{|p{2.5cm}|p{1.65cm}|p{1.65cm}|p{2.5cm}|}
\hline
 & \textbf{Scenario 1} & \textbf{Scenario 2} & \textbf{Scenario 3} \\ \hline
\textbf{Description} & Pressure Indicator \newline Control Failure & Nitrogen Valve Primary Source \newline Failure & Temperature Indicator Control Failure \\ \hline
\textbf{Complexity Level} & Low & Medium & High \\ \hline
\textbf{Alarm Flood} & No & No & Yes \\ \hline
\end{tabular}
\end{table}

\section{Feature Extraction}

From the data obtained from each participant during the experiment, we derived performance and behavioural metrics, as shown in \ref{tab:metrics}. These measures were derived from the raw simulator logs. Subsequently, all measures are used for the comparison of groups/support configurations, while a few behavioural metrics, which can be collected practically in a real control room setup, are used to answer the second research question.

\subsection{Performance and System}

Performance-based metrics obtained in this study include recovery time, accuracy, overall performance, and consequence. The no of alarms activated is a system feature but also tells the outcomes of operators' actions - correct or incorrect. Figure \ref{tab:metrics} shows the definition of these measures for each scenario in the experimental study. As observed, the extraction of some of the metrics for scenario 2 was similar to scenario 1 and varied with scenario 3, given the difference between the scenario goals. This was because scenario 3 was related to an entirely different section of the plant, the reactor and absorber sections, while scenarios 1 ad 2 were related to the Tank section.

Recovery time for scenarios 1 and 2 is extracted based on the feature '\(All2_1\)' (Pressure Alarm Low). In Scenario 3, the recovery time does not count as recovery is meant for the field operator and not the control room operator, given that an issue on the heat recovery (REC3) cooling water sytem was intentionally induced and can only be fixed on the field (see Figure \ref{fig:sim_main}). The measure was derived from the raw data variable 'TMAXREACTORE' which is the reactor temperature. Survival time, however, is used in place of response time for scenario 3. It means the operators' ability to maintain a no shutdown or reactor overheating situation until the end of the test at the 18th minute. The assumption was that at this 18th minute, the field operator should have resolved the issue with the REC3 cooling system. 

Accuracy was derived based the variables '\(FN2serb1O_1\)' (Primary Nitrogen Flow), '\(MWpopOld_1\)' (Manual Pump power value) and '\(Toutrec2c_1\)' (Temperatura Out REC1) for scenarios 1, 2 and 3 respectively.  It refers to the mean squared error (mse) between their adjusted value and the mean of the range as written in the operating or intervention procedure.

In deriving the overall performance class, the individual group distributions of recovery time was considered. Those classified as 'optimal' fell within the 25th percentile of the study population, 'Good' below the 50th percentile and others 'poor'. The consequence is derived from the raw features ' \(PSERB_1\)', '\(TMAXREATORE_1\)' and '\(emmerO_1\)' depending on these conditions in Table 1, leading to one of the consequences; Safe - rank 1,  Impurity of air in the tank atmosphere - rank 2,  Air in the tank atmosphere - rank 3, Shutdown - rank 4, Implosion - rank 5 (for scenarios 1 and 2 only), Reactor Overheated - rank 5 (for scenario 3 only). 

Lastly, the number of alarms activated was based on all alarm features. Each alarm in the plant was recorded with a feature starting with 'All + the number', for example, \(All2_1\) for pressure alarm low (PAL01). When an alarm is activated the value changes from 0 to 1. The feature extraction code checks for every occurrence of 1 during the scenario, also taking into account any other activation after a precious deactivation and sums this up. 

A description of other features are as shown in Table \ref{tab:metrics}.  An example plot showing the extraction of the reaction time and response time for scenario 1 using one of the participant's outcomes is shown in Figures \ref{fig:reactiontime_ex} and \ref{fig:responsetime_ex}, respectively.
 
\begin{table*}
\caption{Behavioural, System and Performance metrics extracted from simulator logs during the experiment; scenario 1, 2 and 3.}
\tiny
\label{tab:metrics}
\begin{tabular}{@{}p{0.1cm}p{2cm}p{4cm}p{4cm}p{4cm}@{}}

\toprule
No. & Metrics & S1 & S2 & S3 \\ \midrule
\rowcolor[HTML]{EFEFEF} 
1 & \begin{tabular}[c]{@{}p{4cm}@{}}Recovery \\ Time\end{tabular} & \begin{tabular}[c]{@{}p{4cm}@{}}The time it takes to recover the pressure alarm PAL01. \\ This time is calculated from the initiation of the fault at 60 seconds after the start of the experiment until the time PAL01 is cleared. Same for everyone.\end{tabular} & Same as S1. & \begin{tabular}[c]{@{}p{4cm}@{}}For scenario 3, the recovery time is not considered since the operator has to maintain the temperature of the reactor stable until the end of the scenario.
\\ \\ \end{tabular} \\
2 & \begin{tabular}[c]{@{}p{4cm}@{}}Reaction \\ Time\end{tabular} & \begin{tabular}[c]{@{}p{4cm}@{}}The time it takes to switch the Nitrogen valve button from Auto to Manual. \\ This time is calculated from the initiation of the fault at 60 seconds after the start of the experiment until the time ‘manual’ is activated.\end{tabular} & Same as S1. & \begin{tabular}[c]{@{}p{4cm}@{}}We have in this case the time it takes to switch the cooling water system of FAL11 (REC3WMO\_1) from Auto to Manual.\end{tabular} \\
\rowcolor[HTML]{EFEFEF} 
3 & \begin{tabular}[c]{@{}p{4cm}@{}}Response \\ Time\end{tabular} & \begin{tabular}[c]{@{}p{4cm}@{}}The time it takes to adjust the nitrogen valve scale to the correct value. \\ This time is calculated from the initiation of the fault at 60 seconds after the start of the experiment until the last adjustment action on the nitrogen valve scale.\end{tabular} & \begin{tabular}[c]{@{}p{4cm}@{}}The time it takes to switch from \\ primary to backup. \\ If not switched at all then label 'error'. \\ This time is calculated from the initiation of the fault at 60 seconds after the start of the experiment.\end{tabular} & \begin{tabular}[c]{@{}p{4cm}@{}}It is the time of last activity before the spike of trend in the temperature cooling system (Toutrec2c\_1).  In scenario 3 it is rather referred to as the survival time (expected to survive up to 1080 seconds which is the task duration). \end{tabular} \\
4 & Accuracy & \begin{tabular}[c]{@{}p{4cm}@{}}refers to the difference between their adjusted value on the nitrogen valve scale and the mean of the range given in the procedure (Mean Squared Error).\end{tabular} & \begin{tabular}[c]{@{}p{4cm}@{}}The difference between their adjusted value on the pump power valve scale (MWpopOld\_1) and the mean of the range given in the procedure (MSE).\end{tabular} & \begin{tabular}[c]{@{}p{4cm}@{}}The difference between their adjusted value on the temperature cooling system scale (Toutrec2c\_1) and the mean of the range given in the procedure (MSE).\end{tabular} \\
\rowcolor[HTML]{EFEFEF} 
5 & \begin{tabular}[c]{@{}p{4cm}@{}}Overall \\ performance\end{tabular} & \begin{tabular}[c]{@{}p{4cm}@{}}
Optimal: Recovery time $\leq$ (25th percentile).\\
Good: 25th percentile \textless {Recovery time} $\leq$ (50th percentile).\\
Poor: Recovery time \textgreater{50th percentile}.\\
 \end{tabular} & \begin{tabular}[c]{@{}p{4cm}@{}}Same as S1\end{tabular} & \begin{tabular}[c]{@{}p{4cm}@{}}
Optimal: Recovery time $\geq$ (900)\\ 
and the consequence is 'Safe'.\\
Good: 850 $\leq$ (Recovery time)\\ \textless{ 900}. \\
Poor: All other cases\end{tabular}\\
6 & Consequence & \begin{tabular}[c]{@{}p{4cm}@{}}• If PSV01 is activated, \\ then Consequence == Safe – (1)\\    \\ • If PSV01 activated, that is   Pressure \textless 0.98000) and Pressure \textgreater{}= 0.975, \\ then Consequence == Impurity of air in the tank atmosphere – (2)\\    \\ Else Consequence == Air in the tank atmosphere – (3)\\    \\ • If PSLL, that is \\ Pressure \textless{}=   0.97000 activated, \\  then Consequence == Shutdown – (4)\\    \\ • OR If 'emmerO\_1, that is \\  Emergency Shutdown == 1, \\ then Consequence == Shutdown – (4)\\    \\ • If all systems failed, that is \\  Pressure \textless{}= 0.97000) and \\ 'emmerO\_1' == 1 \\ then Consequence == Implosion – (5)\end{tabular} & Same as S1. & \begin{tabular}[c]{@{}p{4cm}@{}}If TMAXREATORE\_1, that is \\ Reactor Temperature \textgreater{}= 400 \\    \\ then Consequence == Reactor Overheated - Score: (5)\end{tabular} \\
\rowcolor[HTML]{EFEFEF} 
7 & \begin{tabular}[c]{@{}p{4cm}@{}}Alarms \\ Silenced\end{tabular} & No. of alarms silenced during the scenario. & Same as S1. & Same as S1. \\
8 & Acknowledgement & No. of alarms acknowledged during the scenario. & Same as S1. & Same as S1. \\
\rowcolor[HTML]{EFEFEF} 
9 & Mimics opened & No. of mimics opened throughout the scenario. & Same as S1. & Same as S1. \\
10 & No. \newline of procedures & No. of procedure opened throughout the scenario. & Same as S1. & Same as S1. \\
\rowcolor[HTML]{EFEFEF} 
11 & No. of alarms & No. of alarms activated. & Same as S1. & Same as S1. \\
12 & Error & If at the end of the scenario, the pressure is increasing or has reached a normal level and shutdown is not reached & Same as S1. & If the high high temperature alarm is reached in the reactor. \\ \bottomrule

\end{tabular}
\end{table*}

\subsection{Behavioural Feature Extraction}

\begin{figure}
  \begin{minipage}[b]{0.50\linewidth}
    \centering
    \includegraphics[width=\linewidth]{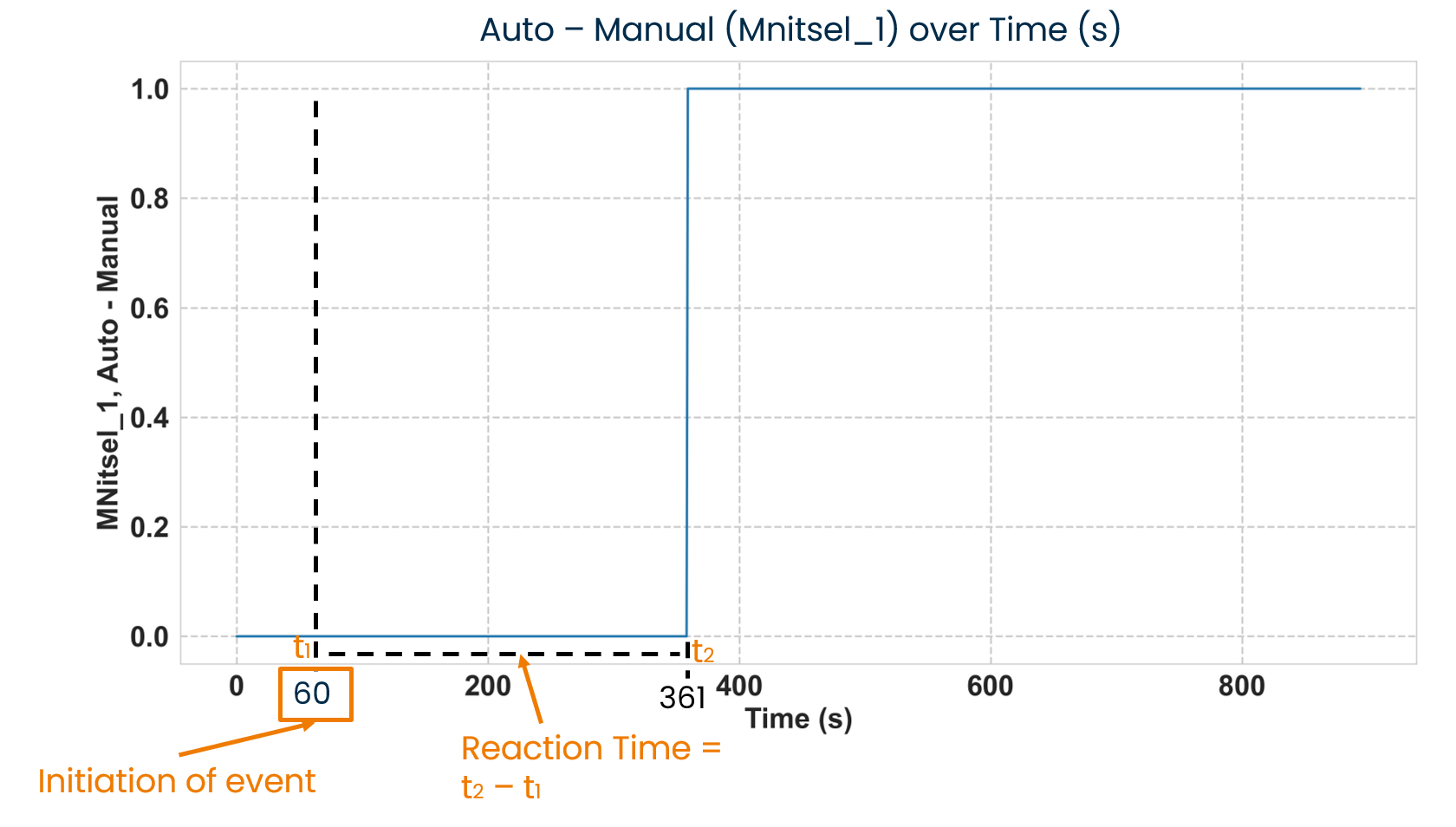}
    \caption{Reaction time, example from Participant 10, Scenario 1}
    \label{fig:reactiontime_ex}
  \end{minipage}
  \hfill
  \begin{minipage}[b]{0.50\linewidth}
    \centering
    \includegraphics[width=\linewidth]{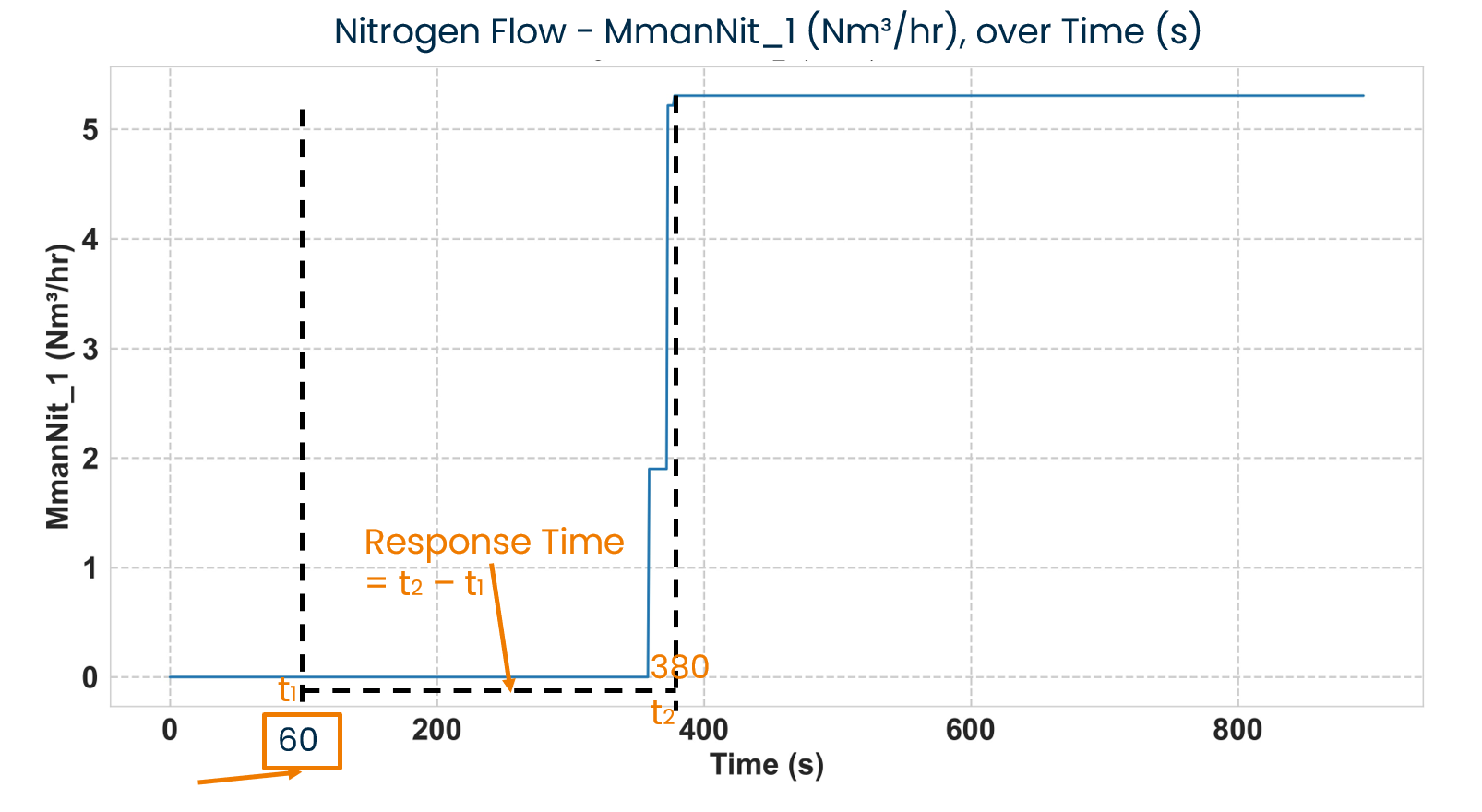}
    \caption{Response Time, example from Participant 10, Scenario 1}
    \label{fig:responsetime_ex}
  \end{minipage}
\end{figure}

The metrics counted as behavioural in this study include reaction time, response time, no procedures opened, no mimics, silenced alarms, and acknowledgement of alarms. These show the operators' behaviour during the task, which is influenced by different underlying factors. They show behaviours that inform if, for example, they have a good situational awareness, such as from the increased opening of mimics when just one mimic is required for the task. Or the acknowledgement of many alarms which show the poor perception of the critical alarm, or the opening of many procedures contrary to expectation which can also be a sign of poor situation awareness. Similar operator-system interaction metrics have been used by \cite{Braarud2021} in classifying workload. 
The reaction time in \ref{fig:reactiontime_ex}is the time it takes to switch the Nitrogen valve button from Auto to Manual or in scenario 3 to switch the REC3 cooling water flow system from manual to automatic. It was calculated based on the raw feature '\(MNitsel_1\)' (Manual Nitrogen Valve Switch) for scenarios 1 and 2, and \(REC3WMO_1\) (Heat Recovery 3 cooling water system Switch) for scenario 3.  

Response time, Figure 2, is the time it takes to adjust the nitrogen valve scale to the correct value. For Scenario 2, it is the time it takes to switch from primary to backup and the time of the last activity before the temperature spikes for Scenario 3. It is based on the features '\(MmanNit_1\)' (Manual Nitrogen Valve), '\(Nitsel_1\)' (Primary to Backup System Control and '\(Toutrec2c_1\)' (Reactor Cooling System) for scenario 1, 2, and 3 respectively.
The number of procedures is based on the feature '\(proclis_1\)' (procedures). The record on the number of procedures were only possible for Group 3 and Group 4 because of the ease with recording the operator system interaction on the display for these groups. The number of mimics/display opened, number of alarms silenced and no of alarms acknowledged was recorded for all groups based on the features '\(intop\)', '\(sAll\)', and '\(AckAll\)', respectively. The detailed description of the simulator log variables from which the metrics were derived is published in \cite{AMAZUDatainBrief2024}

\subsection{Exploring the use of machine learning algorithms}

Several machine learning algorithms have been applied to operator-system interaction data to predict or classify workload \cite{Braarud2021}, \cite{Chen2012}, and situational awareness \cite{Shi2022}, \cite{Yang2021}. Braaud in a study on a Halden facility which is close to process control room operations classified cognitive workload using speech, system data (operator \& crew alarms, acknowledged operator \& crew alarms, human system interface activity and commands) with Random Forest(RF), Support Vector Machine (SVM), Extreme Gradient Boosting (XGB), Neural Networks (NN), Naive Bayes (NB), and k-Nearest Neighbors (KNN) algorithms. For the data used in the crew system, RF, XGB performed better than the others at an accuracy of 61\% and averagely 72\% when trained for a single operator. The important features were the number of alarms and the amount of operator-system interaction. In a different domain, Rahman while classifying cognitive load of non-contact based drivers used Logistic Regression (LR), Linear Discriminant Analysis (LDA), SVM and NN, with LR outperforming the other methods \cite{Rahman2020}. Similarly, while classifying workload using driver performance and psycho-physiological data, Solovey reported that logistic regression and naive bayes classifiers outperformed the others (Decision Trees, Multilayer Perceptron, Nearest neighbour) \cite{Solovey2014}. The algorithms mentioned so far were likewise used for workload or situational awareness classifications by researchers whose inputs were solely from eye-trackers and EEG \cite{Yang2021}, \cite{Iqbal2024}. Though these studies have been on workload and situational awareness predictions and classification, similar algorithms are explored in this work for error prediction specifically Step-wise Logistic Regression (LR) and Tree Augmented Naive Bayes (TAN).

A general recommendation from the reviewed studies has been that using multi-modal inputs improve prediction accuracies than when single modalities are used. Though the focus of this work is on the operational logs, the subjective features for workload, situational awareness, training and experience are later added to observe how they influence outcomes. Also it is important to highlight that the predictions in the above cited studies were carried out based on a range that varies between 13 to 99 subject participants observations.

\subsection{Data Pre-Processing and the choice of suitable algorithms}

The following algorithms are used to address the second research question.

\subsubsection{Step-wise Logistic Regression Analysis}

Step-wise Logistic Regression is a method used for building a predictive model by selecting the most significant variables incrementally until certain criteria, like the model fit or statistical significance, are met. This approach is particularly advantageous when dealing with a large set of potential predictors, as it helps to identify the most relevant ones while excluding those that do not contribute significantly to the model's predictive power.

The logistic regression model predicts the probability of a binary outcome based on one or more predictor variables. The logistic function used in this model is defined as:

\[
P(Y = 1 \mid X) = \frac{1}{1 + e^{-(\beta_0 + \beta_1X_1 + \beta_2X_2 + \ldots + \beta_nX_n)}}
\]

where \(Y\) is the binary dependent variable (e.g., error occurrence), \(X_1, X_2, \ldots, X_n\) are the independent variables (predictors), and \(\beta_0, \beta_1, \ldots, \beta_n\) are the coefficients of the model.

The step-wise approach involves the following steps:

\begin{enumerate}
    \item \textbf{Initialization}: Start with an empty model or a model with a single predictor.
    \item \textbf{Forward Selection}: Add predictors to the model one by one. At each step, add the predictor that improves the model the most, usually based on the Akaike Information Criterion (AIC) or Bayesian Information Criterion (BIC).
    \item \textbf{Backward Elimination}: After adding a new predictor, remove any predictor that does not significantly improve the model. This is also done based on AIC or BIC values.
    \item \textbf{Iteration}: Repeat steps 2 and 3 until no further predictors can be added or removed based on the chosen criterion.
\end{enumerate}

The goal of step-wise logistic regression is to achieve a balance between model complexity and predictive accuracy, thereby avoiding overfitting by including only those predictors that significantly contribute to the model (as observed using the p values) and to the accuracy score. In our study, we applied step-wise logistic regression to identify key factors influencing the likelihood of human error. By iteratively adding and removing variables, we derived a parsimonious model that effectively predicts error occurrence based on these behavioural metrics.

The results from the step-wise logistic regression analysis provide insights into the relative importance of different predictors and help to refine our understanding of the factors contributing to human errors in control room environments.

\subsubsection{Bayesian Network}

In this study, we employ Naive Bayes classifiers, and Tree-Augmented Naive Bayes models to predict operational outcomes. These models are valuable tools for understanding and modeling the complexities of operational data.

\paragraph{Bayesian Networks}

A Bayesian Network is a probabilistic graphical model that represents a set of variables and their conditional dependencies through a directed acyclic graph (DAG). In this graph, nodes represent random variables, while directed edges indicate probabilistic dependencies between these variables. This structure allows BNs to model (causal) interactions and predict events effectively. Each node is associated with a conditional probability distribution that quantifies the effect of the parent nodes on the node itself. This factorization simplifies the joint probability distribution, making complex problems more manageable.

A discrete Bayesian Network $\mathcal{N} = (\mathcal{V}, \mathcal{E}, \mathcal{P})$ is formally defined as follows \cite{jensen2007bayesian}:

\begin{itemize}
  \item $\mathcal{V} = \{v_1, v_2, \ldots, v_n\}$ is a set of nodes, each representing a discrete random variable $X_v$.
  \item $\mathcal{E}$ is a set of directed edges (arcs) between the nodes, forming a DAG. Each edge $(v_i, v_j) \in \mathcal{E}$ indicates a direct probabilistic influence from node $v_i$ to node $v_j$.
  \item $\mathcal{P}$ is a set of conditional probability distributions, one for each variable $X_v$ given its parents $X_{pa(v)}$ in the graph $\mathcal{G}$. Specifically, for each $v \in \mathcal{V}$, there is a conditional probability distribution $P(X_v \mid X_{pa(v)})$.
\end{itemize}

A Bayesian Network encodes the joint probability distribution over a set of random variables $\mathcal{X} = \{X_1, X_2, \ldots, X_n\}$. The joint probability distribution $P(\mathcal{X})$ factors according to the structure of the DAG, leveraging the conditional independencies implied by the graph. This factorization is given by:

\begin{equation}
 P(\mathcal{X}) = \prod_{v \in \mathcal{V}} P(X_v \mid X_{pa(v)})
\end{equation}

This equation illustrates that the joint probability distribution is decomposed into a product of conditional probabilities, each corresponding to a node given its parents in the DAG. This factorization significantly reduces the complexity of representing the joint distribution, especially for large networks.

\paragraph{Naive Bayes}

The Naive Bayes classifier is a simple yet powerful probabilistic model used for classification tasks. It assumes conditional independence between feature variables given the class variable \(C\), which simplifies the computation of the joint probability distribution. Despite this simplification, the Naive Bayes is often surprisingly effective and is robust to noise  \cite{Anders2013}. Let \(\mathbf{x} = (x_1, x_2, \ldots, x_n)\) represent an instance to be classified, and \(C\) be the class variable. The Naive Bayes classifier computes the probability of the instance as follows:

\[
P(x_1, x_2, \ldots, x_n \mid C) = \prod_{i=1}^{n} P(x_i \mid C)
\]

This independence assumption allows the model to scale linearly with the number of features, making it computationally efficient and suitable for large datasets.

\paragraph{Tree-Augmented Naive Bayes (TAN)}

The TAN model enhances the Naive Bayes classifier by introducing dependencies between feature variables. This method improves the model's expressiveness while retaining computational efficiency. The TAN model adds a tree structure over the feature variables, maintaining a balance between simplicity and capturing essential dependencies  \cite{Anders2013}.

Let \(\mathbf{X} = (X_1, X_2, \ldots, X_n)\) represent the set of feature variables, with \(X_1\) as the root variable of the tree structure . The TAN model is built through the following steps:

\begin{itemize}
\item Calculate the conditional mutual information \(I(X_i, X_j|C)\) for all pairs of feature variables.
\item Construct a maximum-weight spanning tree \(T_S\) for the complete graph over \(\mathbf{X} \setminus \{C\}\), using the mutual information as edge weights.
\item Direct the edges in \(T_S\) by selecting a root in \(\mathbf{X} \setminus \{X_1\}\) and orienting edges away from it.
\item Construct the graph corresponding to a Naive Bayes model with \(X_1\) as the root of \(T_S\) and add the directed edges of \(T_S\) to it.
\end{itemize}

By incorporating these dependencies, TAN provides a more accurate representation of the relationships between variables compared to the standard Naive Bayes classifier.

\paragraph{Entropy minimisation algorithm}

In this study, we employed an entropy-based discretization algorithm to categorize continuous variables into discrete intervals. This method, introduced by Fayyad and Irani  \cite{fayyad1993multi}, recursively partitions the range of a continuous attribute, selecting cut points that maximize the information gain ratio. The algorithm aims to find optimal thresholds that preserve the class-attribute interdependence while reducing the data's complexity. This approach is particularly useful in preparing data for Bayesian Network analysis, as it allows for the efficient handling of continuous variables while maintaining their predictive power. The entropy discretization method has been shown to be effective in various machine learning applications, often outperforming other discretization techniques in terms of maintaining classification accuracy \cite{liu2002discretization}.


\section{Results}

Data missing from the general analysis, specifically for accuracy (n = 6) was filled by adding '0'. Else, there were no missing values for the other variables and groups. In total 87 participant data were included in the analysis; 22 participant data from group 1, 23 for group 2, 21 group 3 and 21 group 4. Analyses were performed using Python programming.

\section*{Group-wise Comparisons across Scenarios}

In this study, we conducted a comprehensive statistical analysis to examine the effects of experimental conditions across three distinct scenarios on various operational variables. The normality of data distributions was evaluated using the Shapiro-Wilk test \cite{shapiro1965analysis}, while the Levene test \cite{levene1960robust} was utilized to assess the equality of variances among groups. Based on the results of these preliminary tests, the appropriate statistical tests were applied: the Student's t-test \cite{student1908probable} was used for data exhibiting normal distribution and equal variances, the Welch's t-test \cite{welch1947generalization} for data with normal distribution but unequal variances, and the Wilcoxon Rank-Sum Test \cite{mann1947test}  for data not adhering to normality assumptions. The next sections provide a detailed exposition of the outcomes derived from each scenario, presenting a nuanced understanding of how experimental conditions influenced the observed variables. 

\subsection*{Comparison between Group 1 and Group 2}

In this section, we undertake a comparative analysis of operational variables between two distinct groups to assess the impact of alarm prioritisation on operator performance. Group 1 operates without the benefit of alarm prioritisation, whereas Group 2 utilizes an alarm prioritisation system. Through this comparison, we aim to determine the effectiveness of alarm prioritisation in enhancing operational efficiency and decision-making processes among operators. This analysis is pivotal in determining whether the implementation of prioritized alarms can significantly influence operator response times, accuracy, and overall performance in managing operational tasks.

\begin{description}
    \item[Scenario 1:] \hfill
    \begin{itemize}
        \item \textbf{Acknowledgement} demonstrated a significant difference; G2 had a higher acknowledgment rate (p=0.015).
        \item \textbf{No. of alarms} revealed significant differences; G2 had a lower number of alarms (p=0.033).
        \item The rest of the variables showed no significant differences.
    \end{itemize}
    \item[Scenario 2:] \hfill
    \begin{itemize}
        \item All tested variables show no significant differences.
    \end{itemize}
    \item[Scenario 3:] \hfill
    \begin{itemize}
        \item \textbf{Accuracy}  showed significant differences with G1 demonstrating higher accuracy (p-values: 0.008).
        \item \textbf{Alarms silenced} and \textbf{Acknowledgement} showed significant differences with  G2 having a higher rate of acknowledged and silenced alarms (p-values: 0.004, 0.007).
        \item \textbf{Recovery time}, \textbf{response time} showed significant differences with  G2 having a higher recovery and response time (p-values: 0, 0.002).
        \item \textbf \textbf{Consequence} showing borderline statistical significant difference (p-value 0.054) with G2 having better consequences. 
        \item \textbf{Error rate} showed significant differences with G2 demonstrating better performance (p-values: 0.034) 
        \item The rest of the variables showed no significant differences .

    \end{itemize}
\end{description}

The comparison between Group 1 and Group 2 highlights the influence of alarm prioritization on operator performance. In Scenarios 1 and 2, where the number of alarms was relatively low, the differences between the groups were minimal. However, Group 2 consistently showed a higher acknowledgment rate and fewer alarms in Scenario 1, suggesting better alarm management.

In Scenario 3, significant differences emerged, with Group 2 exhibiting higher acknowledgment and silencing rates of alarms, better overall performance, and better outcomes in terms of consequences. Despite Group 2 having higher response times, these findings suggest that alarm prioritization can enhance operator performance and efficiency, particularly in more alarm-intensive scenarios. This underscores the potential benefits of implementing alarm prioritization systems to improve operational tasks, though the specific context and number of alarms play a crucial role in the observed outcomes.

\begin{table}[H]
\caption{Statistical test result of the comparison between Group 1 and Group 2.}
\centering
\scriptsize
\begin{tabular}{@{}lllSSSS@{}}
\toprule
Variable          & Scenario  & Test Used               & {Shapiro-Wilk G1} & {Shapiro-Wilk G2} & {Levene Test} & {p-value} \\ \midrule
Accuracy & S1 & Wilcoxon Rank-Sum & 0 & 0 & 0.807 & 0.751 \\
Accuracy & S2 & Wilcoxon Rank-Sum & 0 & 0 & 0.331 & 0.364 \\
Accuracy & S3 & Wilcoxon Rank-Sum & 0 & 0 & 0.263 & \textbf{0.008} \\
Acknowledgement & S1 & Wilcoxon Rank-Sum & 0 & 0 & 0.441 & \textbf{0.015} \\
Acknowledgement & S2 & Wilcoxon Rank-Sum & 0.015 & 0.008 & 0.833 & 0.794 \\
Acknowledgement & S3 & Wilcoxon Rank-Sum & 0 & 0.008 & 0.348 & \textbf{0.007} \\
Alarms silenced & S1 & Wilcoxon Rank-Sum & 0 & 0 & 0.156 & 0.442 \\
Alarms silenced & S2 & Wilcoxon Rank-Sum & 0 & 0 & 0.761 & 0.23 \\
Alarms silenced & S3 & Wilcoxon Rank-Sum & 0 & 0.206 & 0.689 & \textbf{0.004} \\
Consequence & S1 & Chi-Squared &  & &  & 0.469 \\
Consequence & S2 & Chi-Squared &  & & & 0.985 \\
Consequence & S3 & Chi-Squared &  & & & 0.054 \\
Mimics opened & S1 & Wilcoxon Rank-Sum & 0 & 0.003 & 0.916 & 0.385 \\
Mimics opened & S2 & Wilcoxon Rank-Sum & 0 & 0.008 & 0.911 & 0.96 \\
Mimics opened & S3 & Wilcoxon Rank-Sum & 0.066 & 0 & 0.768 & 0.713 \\
No. of alarms & S1 & Wilcoxon Rank-Sum & 0 & 0 & 0.302 & \textbf{0.033} \\
No. of alarms & S2 & Wilcoxon Rank-Sum & 0 & 0 & 0.593 & 0.567 \\
No. of alarms & S3 & t-test & 0.054 & 0.329 & 0.929 & 0.584 \\
Overall performance & S1 & Chi-Squared & & & & 0.566 \\
Overall performance & S2 & Chi-Squared & && & 0.648 \\
Reaction time & S1 & Wilcoxon Rank-Sum & 0 & 0.002 & 0.868 & 0.136 \\
Reaction time & S2 & Wilcoxon Rank-Sum & 0.009 & 0.026 & 0.389 & 0.481 \\
Reaction time & S3 & t-test & 0.465 & 0.135 & 0.439 & 0.636 \\
Recovery time & S1 & Wilcoxon Rank-Sum & 0.004 & 0.656 & 0.493 & 0.162 \\
Recovery time & S2 & t-test & 0.946 & 0.665 & 0.841 & 0.352 \\
Response time & S1 & Wilcoxon Rank-Sum & 0 & 0.022 & 0.861 & 0.622 \\
Response time & S2 & t-test & 0.064 & 0.221 & 0.741 & 0.302 \\
Response time & S3 & Wilcoxon Rank-Sum & 0.012 & 0.002 & 0.881 & \textbf{0.002} \\
Error rate & S1 & Chi-Squared & & & & 1 \\
Error rate & S2 & Chi-Squared & & & & 0.806 \\
Error rate & S3 & Chi-Squared & & & & \textbf{0.034} \\
\end{tabular}
\label{tab:stat_test_G1_G2}
\end{table}

\subsection*{Comparison between Group 2 and Group 3}

In this analysis, we compare the performance impacts of transitioning from paper-based to digital procedure management between two distinct experimental groups. Group 2 retains the use of paper-based procedures alongside alarm prioritization, whereas Group 3 employs an on-screen procedure display, maintaining the same alarm prioritization system. This comparison aims to explore the effectiveness of digital procedures in enhancing operator interaction, reducing response times, and improving the accuracy and efficiency of decision-making processes. The results from this section will provide insights into whether the digitalization of procedures significantly benefits operators in terms of operational performance and task management.

\begin{description}
    \item[Scenario 1 (S1):] \hfill
    \begin{itemize}
        \item \textbf{No. of alarms} approached significance, suggesting possible differences (p=0.058) with Group 3 having a higher number of alarms.
        \item \textbf{Response time} showed significant differences with G3 having a lower response time (p=0.029).
        \item Other variables showed no significant differences.
    \end{itemize}
    \item[Scenario 2 (S2):] \hfill
    \begin{itemize}
        \item Only \textbf{Overall performance} showed significant differences with G3 having more good performance and lower poor performance but G2 having more Optimal but also poor performance
    \end{itemize}
    \item[Scenario 3 (S3):] \hfill
    \begin{itemize}
        \item \textbf{Acknowledgement}  showed significant differences with G3 demonstrating a lower alarm acknowledgment  (p-values: 0.005).
        \item Other variables showed significant differences.
    \end{itemize}
\end{description}

\begin{table}[H]
\caption{Statistical test result of the comparison between Group 2 and Group 3.}
\centering
\scriptsize
\begin{tabular}{@{}lllSSSS@{}}
\toprule
Variable          & Scenario  & Test Used               & {Shapiro-Wilk G1} & {Shapiro-Wilk G2} & {Levene Test} & {p-value} \\ \midrule

Accuracy          & S1       & Wilcoxon Rank-Sum & 0     & 0     & 0.912 & 0.162 \\
Accuracy          & S2       & Wilcoxon Rank-Sum  & 0     & 0     & 0.313 & 0.601 \\
Accuracy          & S3       & Wilcoxon Rank-Sum & 0     & 0     & 0.557 & 0.758 \\
Acknowledgement      & S1       & Wilcoxon Rank-Sum & 0     & 0.004 & 0.415 & 0.96  \\
Acknowledgement      & S2       & Wilcoxon Rank-Sum & 0.008 & 0.006 & 0.885 & 0.855 \\
Acknowledgement      & S3       & Wilcoxon Rank-Sum  & 0.008 & 0     & 0.169 & \textbf{0.005} \\
Alarms silenced  & S1       & Wilcoxon Rank-Sum   & 0     & 0     & 0.329 & 0.945 \\
Alarms silenced  & S2       & Wilcoxon Rank-Sum  & 0     & 0     & 0.379 & 0.595 \\
Alarms silenced  & S3       & Wilcoxon Rank-Sum  & 0.206 & 0.012 & 0.841 & 0.675 \\
Consequence       & S1       & Chi-Squared         &       &       &       & 1     \\
Consequence       & S2       & Chi-Squared         &       &       &       & 0.383 \\
Consequence       & S3       & Chi-Squared         &       &       &       & 0.664 \\
Mimics opened    & S1       & Wilcoxon Rank-Sum  & 0.003 & 0.018 & 0.54  & 0.576 \\
Mimics opened    & S2       & Wilcoxon Rank-Sum  & 0.008 & 0.003 & 0.359 & 0.963 \\
Mimics opened    & S3       & Wilcoxon Rank-Sum & 0     & 0     & 0.9   & 0.866 \\
No. of alarms    & S1       & Wilcoxon Rank-Sum  & 0     & 0     & 0.183 & 0.058 \\
No. of alarms    & S2       & Wilcoxon Rank-Sum  & 0     & 0     & 0.259 & 0.606 \\
No. of alarms    & S3       & t-test                  & 0.329 & 0.374 & 0.571 & 0.691 \\
Overall performance & S1   & Chi-Squared        &       &       &       & 0.221 \\
Overall performance & S2   & Chi-Squared       &       &       &       & \textbf{0.045} \\
Reaction time       & S1       & Wilcoxon Rank-Sum & 0.002 & 0     & 0.117 & 0.117 \\
Reaction time       & S2       & Wilcoxon Rank-Sum & 0.026 & 0.002 & 0.18  & 0.725 \\
Reaction time       & S3       & Wilcoxon Rank-Sum & 0.135 & 0.04  & 0.283 & 0.085 \\
Recovery time       & S1       & Wilcoxon Rank-Sum  & 0.656 & 0.019 & 0.106 & 0.125 \\
Recovery time       & S2       & t-test                  & 0.665 & 0.19  & 0.1   & 0.466 \\
Response time       & S1       & Wilcoxon Rank-Sum & 0.022 & 0     & 0.451 & \textbf{0.029} \\
Response time       & S2       & Wilcoxon Rank-Sum  & 0.221 & 0.055 & 0.05  & 0.617 \\
Response time       & S3       & Wilcoxon Rank-Sum   & 0.002 & 0.015 & 0.603 & 0.239 \\
Error rate & S1 & Chi-Squared & & & & 1 \\
Error rate & S2 & Chi-Squared & & & & 0.339 \\
Error rate & S3 & Chi-Squared & & & & 0.555 \\
\end{tabular}
\label{tab:stat_test_G2_G3}
\end{table}

Scenario 1 revealed a nearly significant reduction in the number of alarms for Group 3 (p=0.058), suggesting that digital procedures may help in reducing alarm clutter, thereby potentially decreasing the cognitive load on operators. Significantly lower response times in Group 3 (p=0.029) further underscore the efficiency brought about by digital procedures, which likely facilitate quicker access to necessary information and streamline operator interactions.

Scenario 2 highlighted that while Group 3 exhibited more consistently good performance, Group 2 showed a dichotomy of having more optimal but also more poor performances. This indicates that while paper-based procedures can sometimes lead to high performance, perhaps due to familiarity, they also pose risks of significant variability in operator effectiveness, which digital procedures seem to mitigate by providing a more consistent operational framework.

In Scenario 3, Group 3's lower alarm acknowledgment rate and a similar number of alarms suggest less efficient management of alarms (p-values: 0.005). 

Overall, transitioning from paper-based to digital procedures appears to have some positive impact. The findings suggest that digital procedures improve the response time, and reduce the number of alarms in low-complexity scenarios. The limited number of statistically significant differences across the various metrics, especially in complex scenarios, suggests that the impact of digitalizing procedures may be less profound than initially anticipated.

\subsection*{Comparison between Group 3 and Group 4}

This section delves into the comparative analysis of operational variables between two technologically advanced groups to assess the added impact of AI-based decision support. Group 3 operates with alarm prioritization and digital procedures, while Group 4 enhances this setup with AI-driven DSS tools. The DSS detects the fault early and tells the operators about the potential consequences of the fault and the steps to recover. Those steps are similar to the traditional procedure but drastically reduce and adapt to the current situation. By examining these configurations, we aim to identify the incremental benefits provided by the DSS in terms of enhancing operational safety, reducing workload, and improving decision-making accuracy. This comparison is crucial for understanding if the integration of the DSS can meaningfully elevate operator performance and optimize process management under complex operational scenarios.

\begin{description}
    \item[Scenario 1:] \hfill
    \begin{itemize}
        \item \textbf{Acknowledgement}, \textbf{Alarms silenced} and \textbf{No. of alarms} showed significant differences, with G4 having a lower rate of silenced, acknowledged alarms and also a lower number of alarms (p-values: 0.026, 0.01, 0).
        \item \textbf{Overall performance} showed significant differences, with G4 exhibiting superior performance (p-values: 0.004).
        \item \textbf{Reaction time} and \textbf{Recovery time} showed significant differences, with G4 exhibiting lower reaction and recovery time (p-values: 0, 0).
        \item \textbf{No of procedures} showed significant differences, with G4 opening fewer procedures (p-values: 0).
    \end{itemize}
    \item[Scenario 2:] \hfill
    \begin{itemize}
        \item \textbf{Accuracy} and \textbf{No of alarms} showed significant differences with G4 displaying higher accuracy and fewer alarms (p-values: 0.001, 0.008).
        \item \textbf{Overall performance} also showed significant differences with G4 having a superior Overall performance (p=0).
        \item \textbf{Reaction time}, \textbf{Recovery time} and, \textbf{Response time} showed significant differences, with G4 exhibiting lower reaction and recovery time (p-values: 0, 0, 0).
    \end{itemize}
    \item[Scenario 3:] \hfill
    \begin{itemize}
        \item \textbf{Acknowledgement} indicated G4 had a higher acknowledgment rate (p=0.018).
    \end{itemize}
\end{description}

The comparative analysis of Group 3 and Group 4 across various operational scenarios highlights the tangible benefits of integrating AI-based decision support systems in enhancing operational efficiency and safety. The data clearly demonstrates that Group 4, equipped with AI DSS, consistently outperformed Group 3, which utilized only alarm prioritization and digital procedures.

The improved accuracy observed in Scenario 2 for G4 suggests that the DSS helps operators make more precise decisions. The lower number of acknowledged and silenced alarms in Scenario 1, combined with the overall reduction in alarm counts, implies that the DSS enables early detection and resolution of issues, thereby preventing alarms from escalating. This early intervention is further supported by the reduced reaction and response times in Scenarios 1 and 2, indicating that the DSS provides timely and relevant information, allowing operators to act more quickly and effectively.

The consistently lower recovery times across all scenarios for G4 highlight the DSS's role in facilitating faster resolution of problems, thus minimizing downtime and enhancing operational efficiency. Additionally, the fewer procedures opened in Scenario 1 suggest that the DSS offers more streamlined and tailored guidance, reducing the need for extensive procedure consultation.

The improved consequences in Scenario 2 and enhanced overall performance in Scenarios 1 and 2 demonstrate some benefits of the Decision Support System (DSS) in assisting operators. These results suggest that advanced decision support tools can be valuable in control room environments, potentially enhancing decision-making and improving operational performance in certain situations. However, the findings from Scenario 3 reveal limitations of the DSS. In this complex scenario, Group 4 only showed a higher rate of alarm acknowledgment, with no significant impact on other variables. This indicates that during highly challenging situations, the DSS's effectiveness may be limited, highlighting the importance of not over-relying on such systems and the need for further development to address complex operational scenarios.

\begin{table}[H]
\caption{Statistical test result of the comparison between Group 3 and Group 4.}
\centering
\scriptsize
\begin{tabular}{@{}lllSSSS@{}}
\toprule
Variable          & Scenario  & Test Used               & {Shapiro-Wilk G1} & {Shapiro-Wilk G2} & {Levene Test} & {p-value} \\ \midrule

Accuracy & S1 & Wilcoxon Rank-Sum & 0 & 0 & 0.029 & 0.202 \\
Accuracy & S2 & Wilcoxon Rank-Sum & 0 & 0 & 0 & \textbf{0.001} \\
Accuracy & S3 & Wilcoxon Rank-Sum & 0 & 0 & 0.92 & 0.629 \\
Acknowledgement & S1 & Wilcoxon Rank-Sum & 0.004 & 0 & 0.097 & \textbf{0.026} \\
Acknowledgement & S2 & Wilcoxon Rank-Sum & 0.006 & 0.006 & 0.701 & 0.512 \\
Acknowledgement & S3 & Wilcoxon Rank-Sum & 0 & 0.094 & 0.002 & \textbf{0.018} \\
Alarms silenced & S1 & Wilcoxon Rank-Sum & 0 & 0 & 0.152 & \textbf{0.01} \\
Alarms silenced & S2 & Wilcoxon Rank-Sum & 0 & 0 & 0.59 & 0.107 \\
Alarms silenced & S3 & Wilcoxon Rank-Sum & 0.012 & 0.085 & 0.253 & 0.732 \\
Consequence & S1 & Chi-Squared Test & & & & 1 \\
Consequence & S2 & Chi-Squared Test & & & & \textbf{0.004} \\
Consequence & S3 & Chi-Squared Test & & & & 0.276 \\
Mimics opened & S1 & Wilcoxon Rank-Sum & 0.018 & 0.001 & 0.213 & 0.44 \\
Mimics opened & S2 & Wilcoxon Rank-Sum & 0.003 & 0 & 0.006 & 0.201 \\
Mimics opened & S3  & Wilcoxon Rank-Sum & 0 & 0 & 0.252 & 0.268 \\
No. of alarms & S1 & Wilcoxon Rank-Sum & 0 & 0 & 0.581 & \textbf{0} \\
No. of alarms & S2 & Wilcoxon Rank-Sum Test & 0 & 0 & 0.017 & \textbf{0.008} \\
No. of alarms & S3 & t-test & 0.374 & 0.493 & 0.365 & 0.386 \\
Overall performance & S1 & Chi-Squared & & & & \textbf{0.004} \\
Overall performance & S2  & Chi-Squared & & & & \textbf{0} \\
Reaction time & S1 & Wilcoxon Rank-Sum & 0 & 0 & 0.063 & \textbf{0} \\
Reaction time & S2 & Wilcoxon Rank-Sum & 0.002 & 0 & 0.568 & \textbf{0} \\
Reaction time & S3 & Wilcoxon Rank-Sum & 0.04 & 0 & 0 & 0.153 \\
Recovery time & S1 & Wilcoxon Rank-Sum & 0.019 & 0 & 0.836 & \textbf{0} \\
Recovery time & S2 & Wilcoxon Rank-Sum & 0.19 & 0 & 0.54 & \textbf{0} \\
Response time & S1 & Wilcoxon Rank-Sum & 0 & 0.141 & 0.063 & 0.075 \\
Response time & S2 & Wilcoxon Rank-Sum & 0.055 & 0 & 0.324 & \textbf{0} \\
Response time & S3 & Wilcoxon Rank-Sum & 0.015 & 0.157 & 0.762 & 0.06 \\
No. of procedures & S1 & Wilcoxon Rank-Sum & 0 & 0 & 0.358 & \textbf{0} \\
No. of procedures & S2 & Wilcoxon Rank-Sum & 0 & 0 & 0.296 & 0.535 \\
No. of procedures & S3 & Wilcoxon Rank-Sum & 0.015 & 0.026 & 0.192 & 0.136 \\
Error rate & S1 & Chi-Squared & & & & 1 \\
Error rate & S2 & Chi-Squared & & & 1 \\
Error rate & S3 & Chi-Squared & & & & 0.176 \\
\end{tabular}
\label{tab:stat_test_G3_G4}
\end{table}

\begin{figure}[H]
    \centering
    \begin{subfigure}{.45\textwidth}
        \centering
        \includegraphics[scale=0.13]{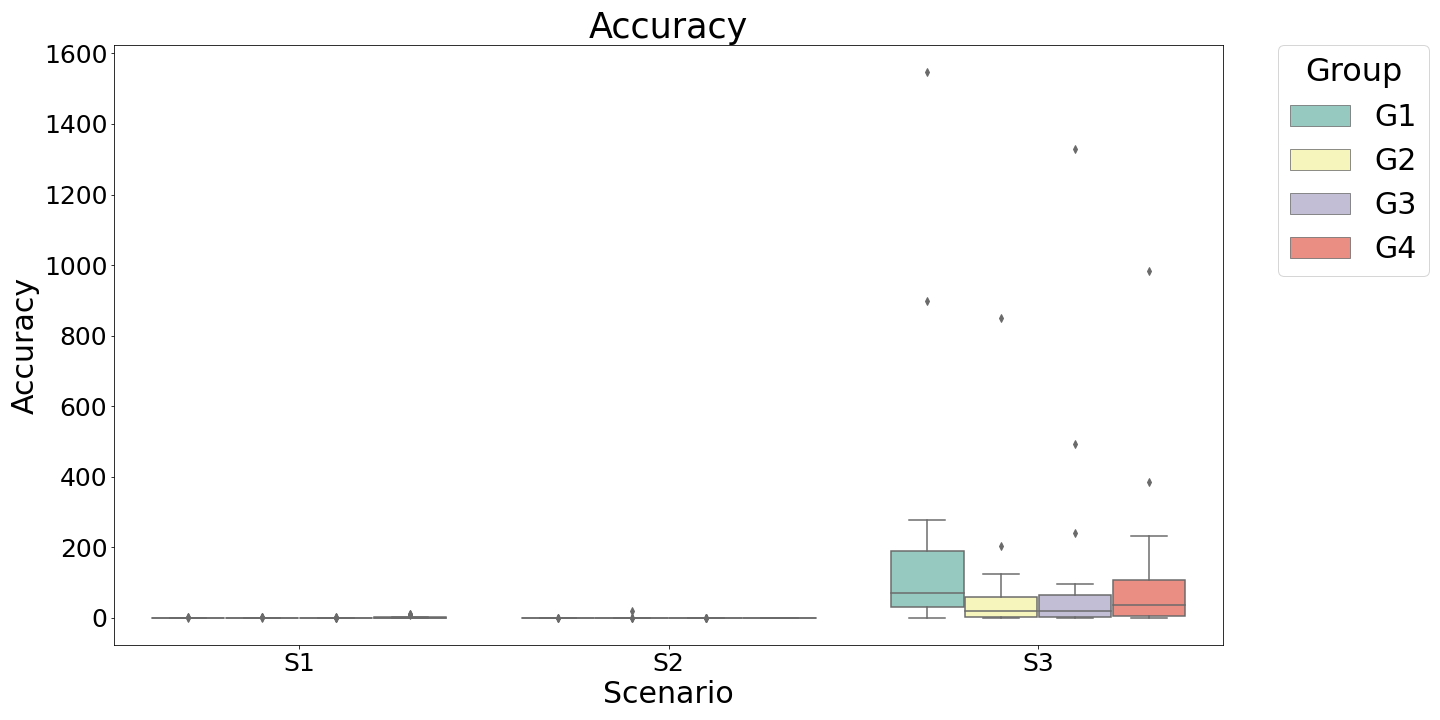}
        \caption{Boxplot of Accuracy}
        \label{fig:Accuracy}
    \end{subfigure}%
    \begin{subfigure}{.45\textwidth}
        \centering
        \includegraphics[scale=0.13]{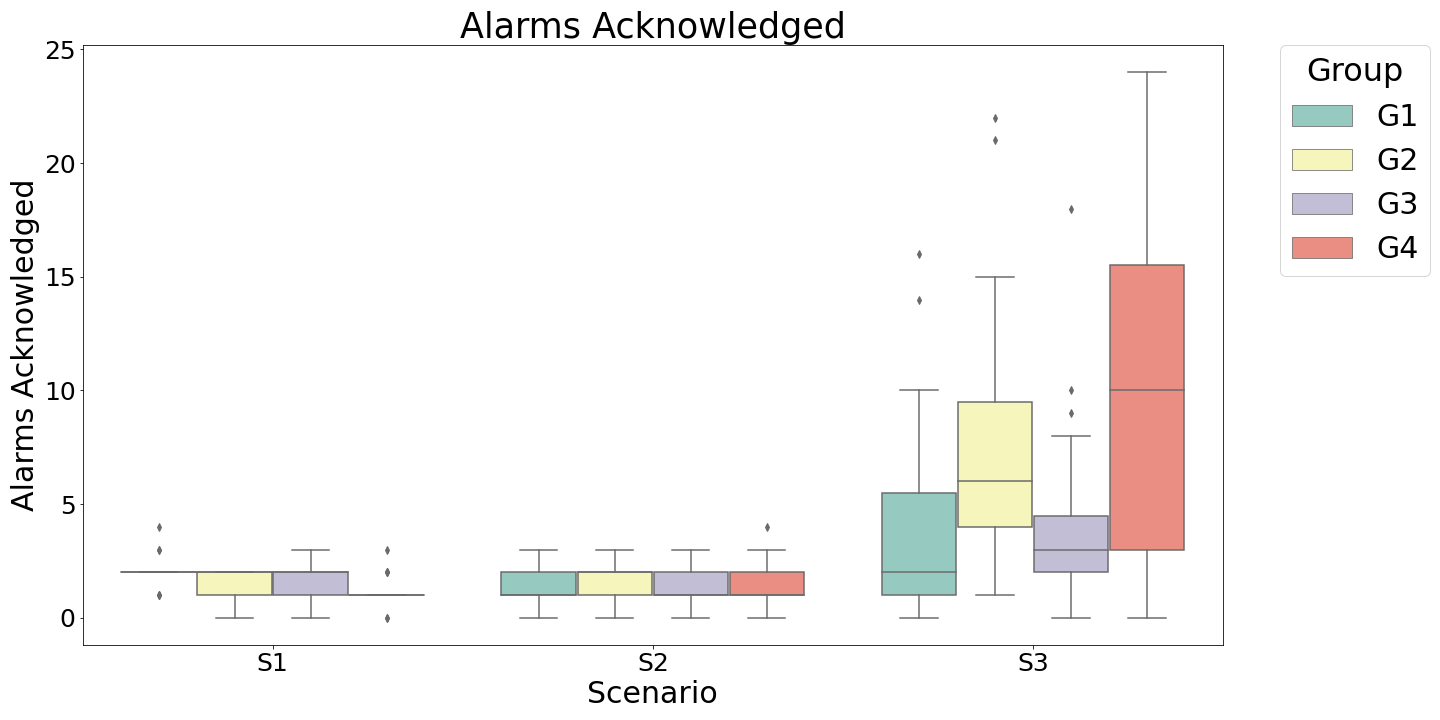}
        \caption{Alarms acknowledged}
        \label{fig:Ack}
    \end{subfigure}
    \caption{Accuracy and Alarms Acknowledged}
\end{figure}

\begin{figure}[H]
    \centering
    \begin{subfigure}{.45\textwidth}
        \centering
        \includegraphics[scale=0.13]{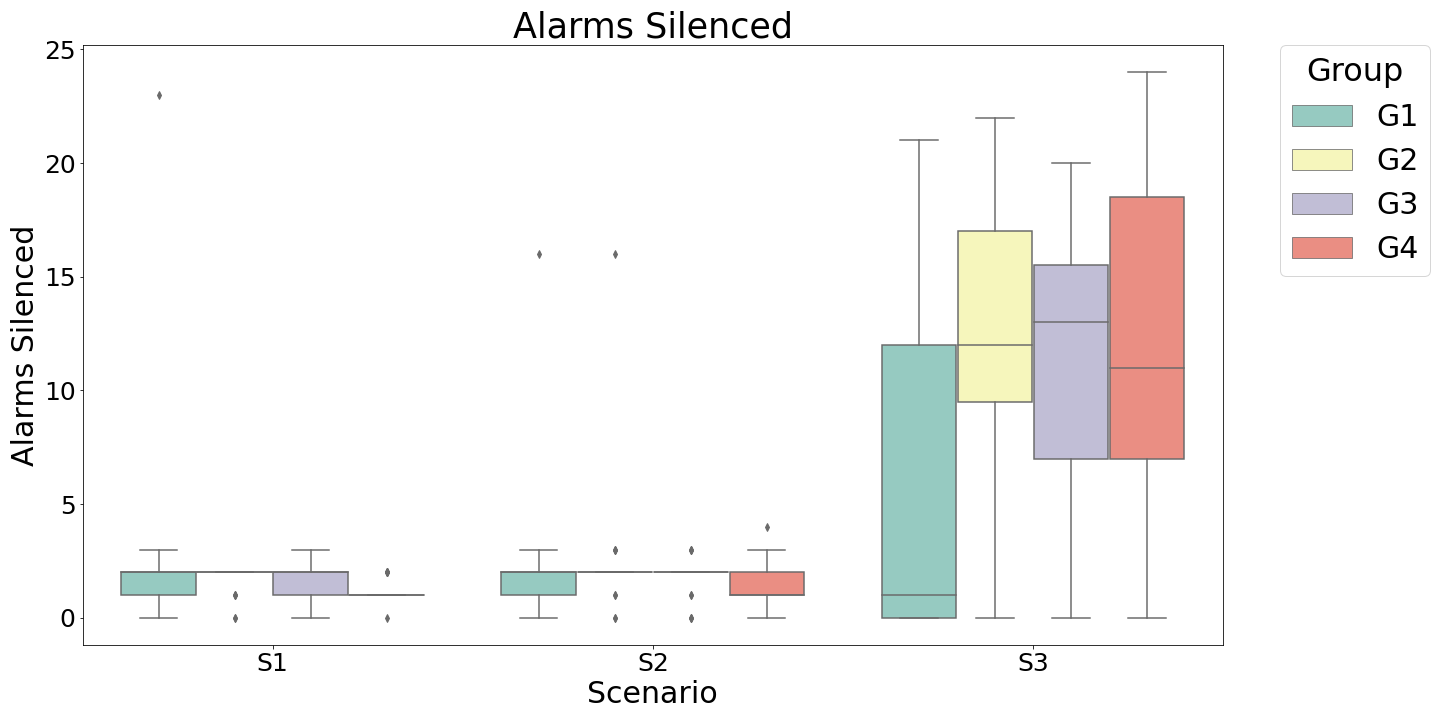}
        \caption{Alarms silenced}
        \label{fig:AlarmsSilenced}
    \end{subfigure}%
    \begin{subfigure}{.45\textwidth}
        \centering
        \includegraphics[scale=0.13]{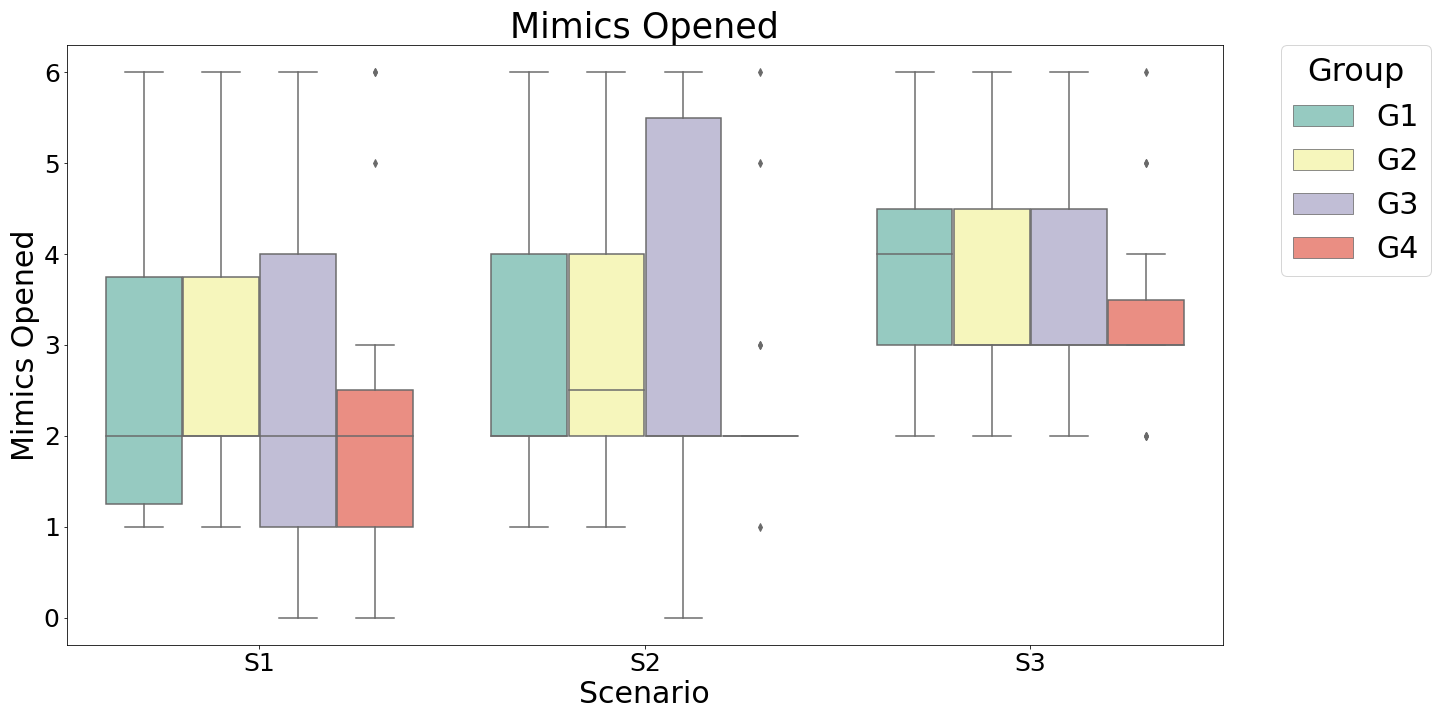}
        \caption{Boxplot of Mimics opened}
        \label{fig:mimic}
    \end{subfigure}
    \caption{Alarms Silenced and Mimics Opened}
\end{figure}

\begin{figure}[H]
    \centering
    \begin{subfigure}{.45\textwidth}
        \centering
        \includegraphics[scale=0.13]{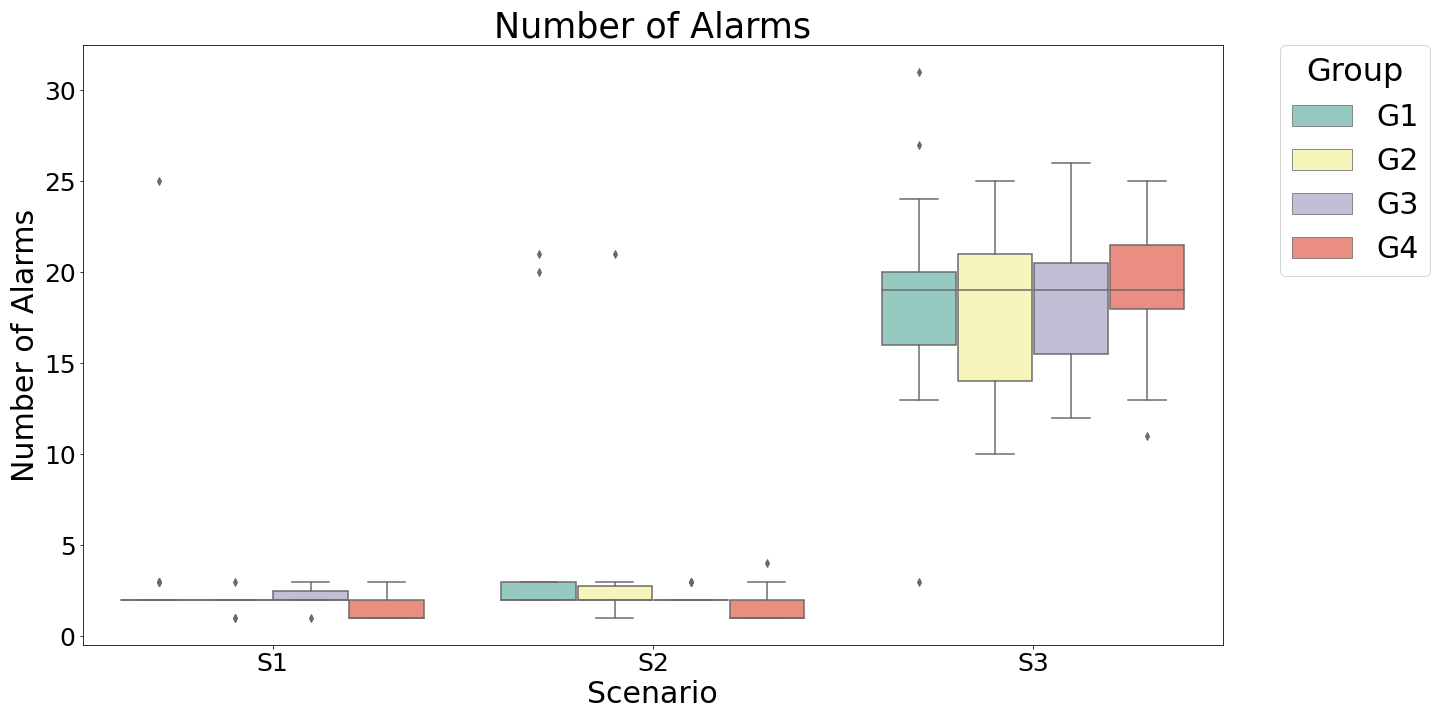}
        \caption{Boxplot of Number of Alarms}
        \label{fig:alarm}
    \end{subfigure}%
    \begin{subfigure}{.45\textwidth}
        \centering
        \includegraphics[scale=0.13]{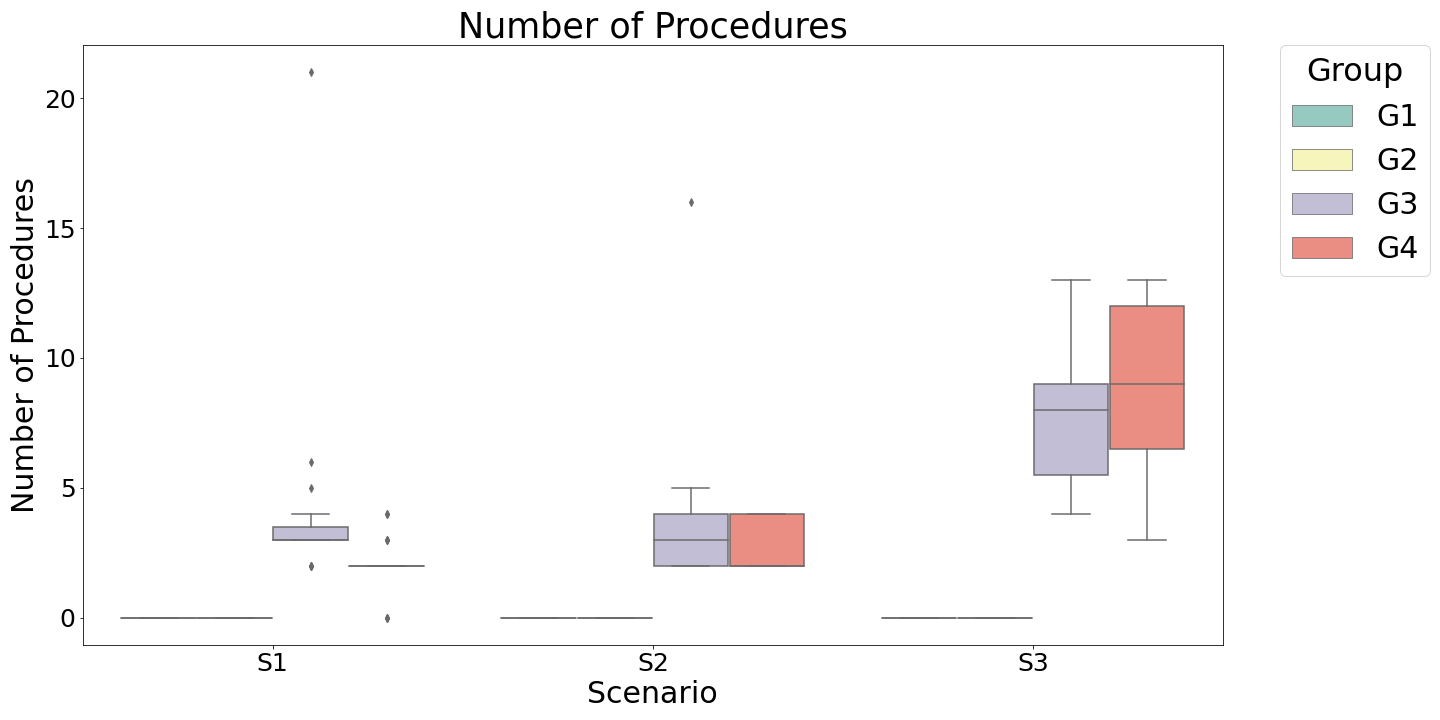}
        \caption{Boxplot of procedure open}
        \label{fig:Procedures}
    \end{subfigure}
    \caption{Number of Alarms and Procedures}
\end{figure}

\begin{figure}[H]
    \centering
    \begin{subfigure}{.45\textwidth}
        \centering
        \includegraphics[scale=0.13]{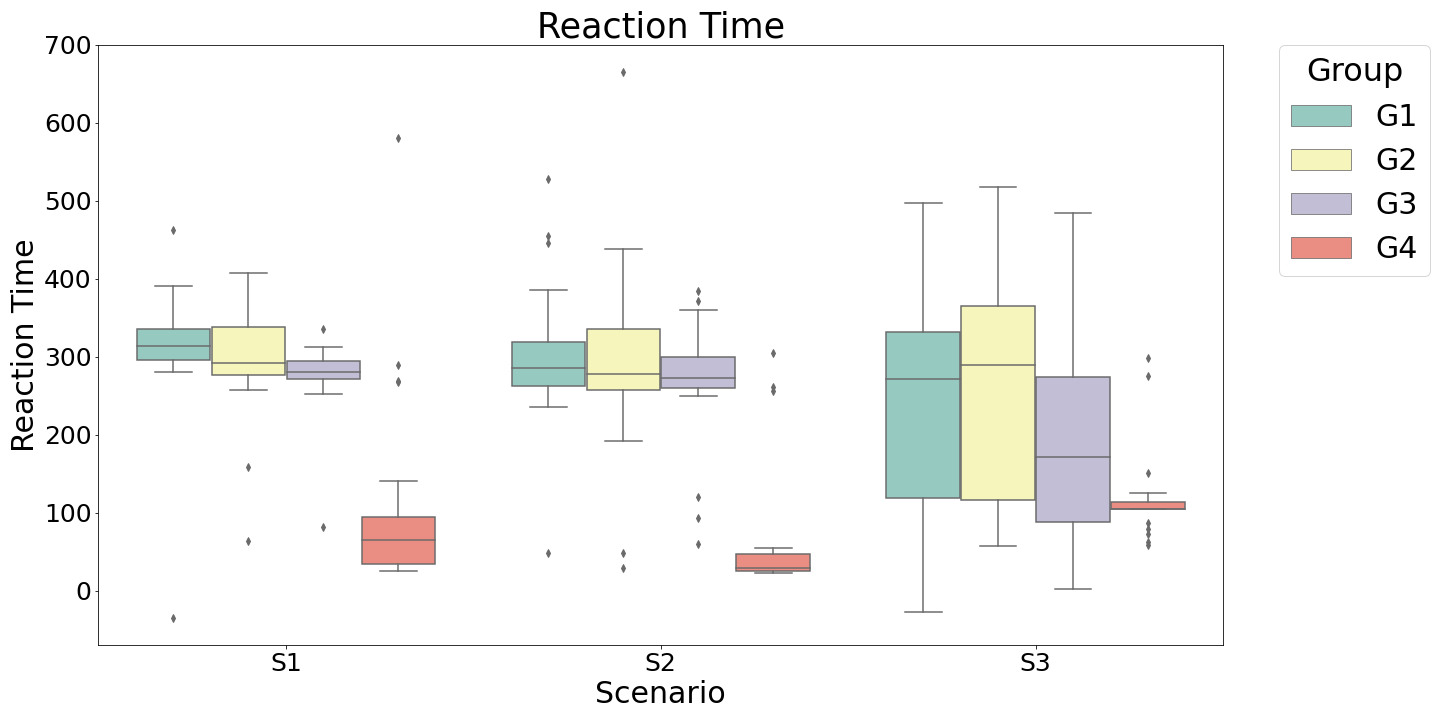}
        \caption{Boxplot of Reaction Time}
        \label{fig:ReactionTime}
    \end{subfigure}%
    \begin{subfigure}{.45\textwidth}
        \centering
        \includegraphics[scale=0.13]{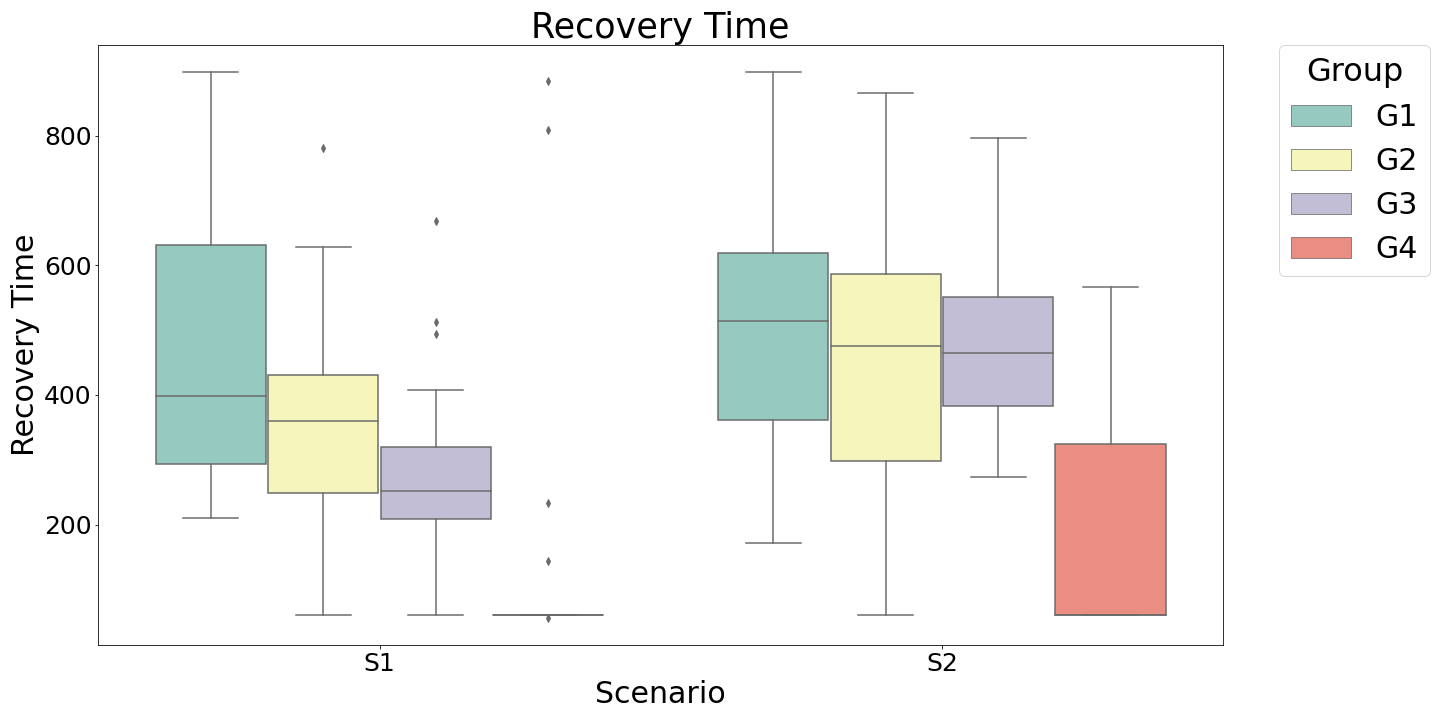}
        \caption{Boxplot of Recovery time}
        \label{fig:recovery}
    \end{subfigure}
    \caption{Reaction and Recovery Times}
\end{figure}

\begin{figure}[H]
    \centering
    \begin{subfigure}{.45\textwidth}
        \centering
        \includegraphics[scale=0.13]{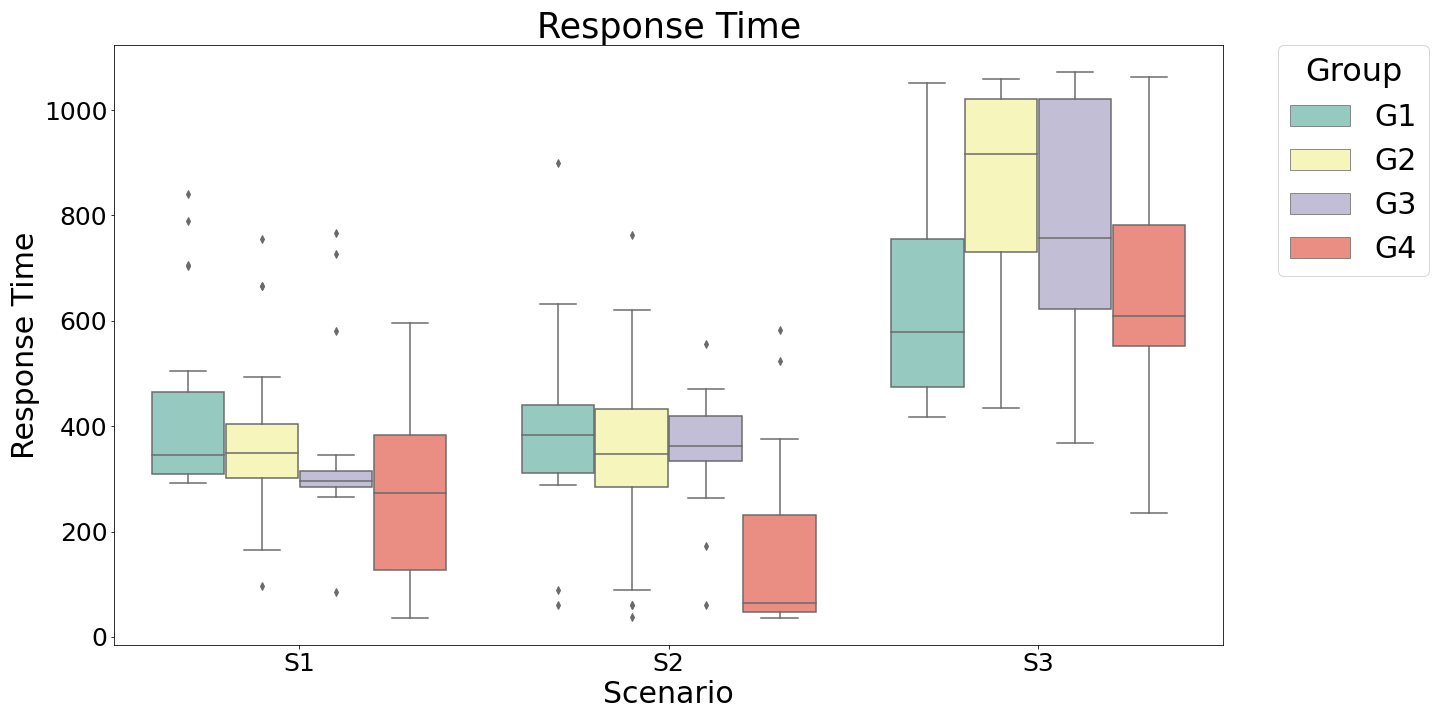}
        \caption{Boxplot of Response time}
        \label{fig:response}
    \end{subfigure}%
    \begin{subfigure}{.45\textwidth}
        \centering
        \includegraphics[scale=0.13]{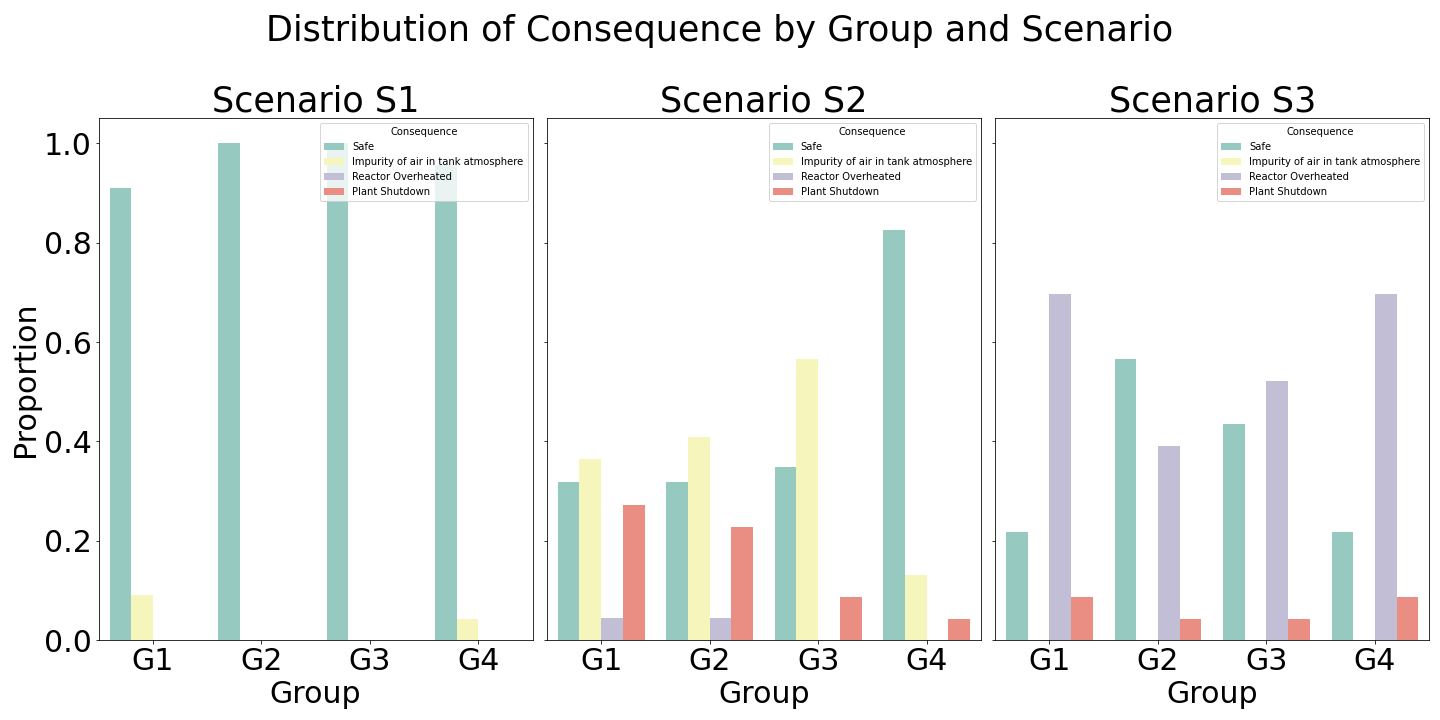}
        \caption{Consequence}
        \label{fig:consequence}
    \end{subfigure}
    \caption{Response Time and Consequence Distribution}
\end{figure}

\begin{figure}[H]
    \centering
    \begin{subfigure}{.45\textwidth}
        \centering
        \includegraphics[scale=0.13]{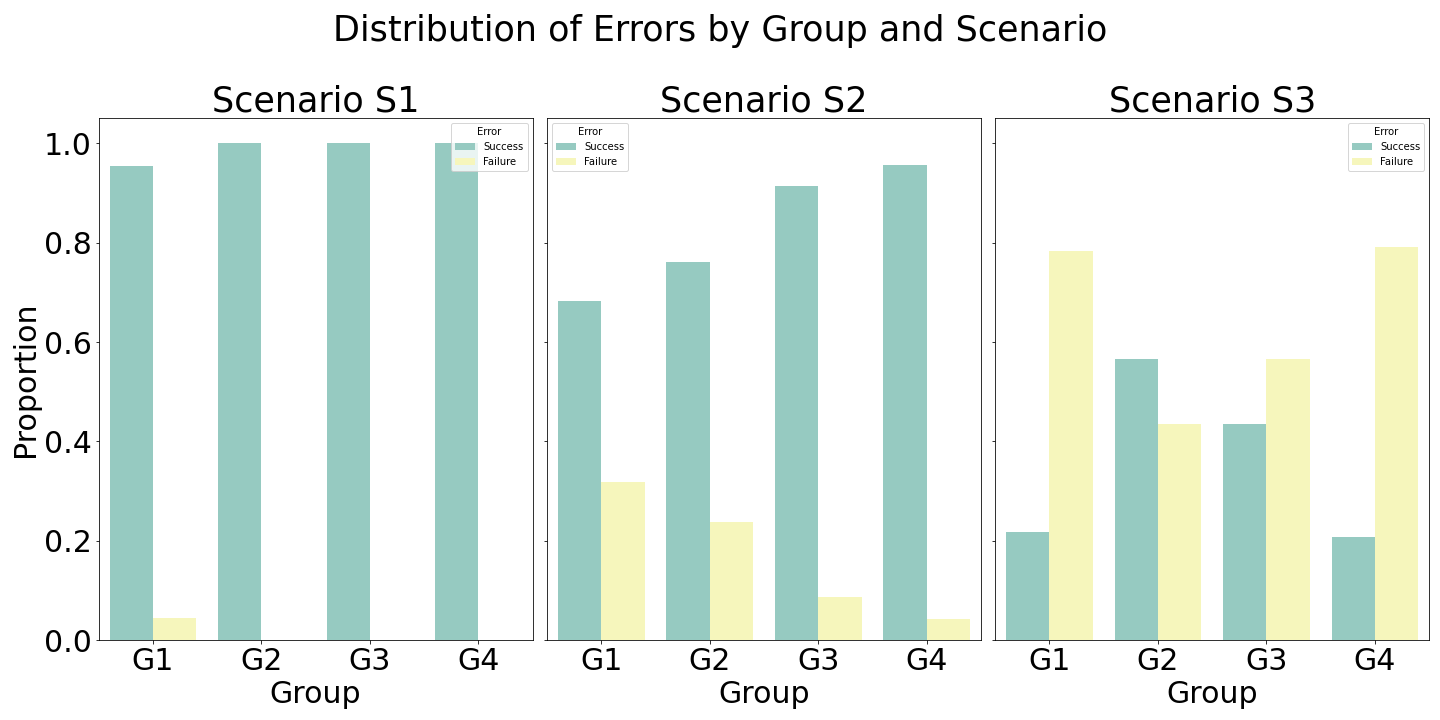}
        \caption{Error rate}
        \label{fig:ErrorRate}
    \end{subfigure}%
    \begin{subfigure}{.45\textwidth}
        \centering
        \includegraphics[scale=0.13]{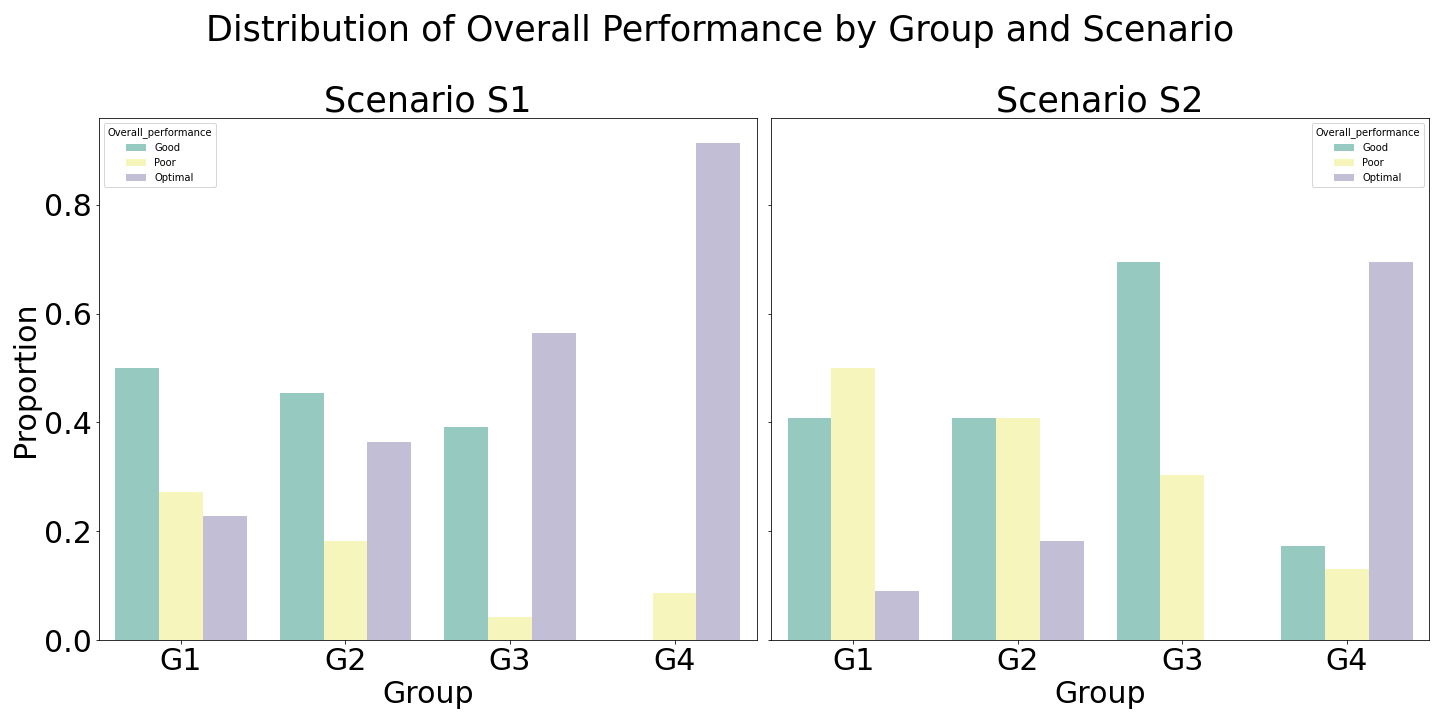}
        \caption{Overall performance}
        \label{fig:OverallPerformance}
    \end{subfigure}
    \caption{Error Rate and Overall Performance Distributions}
\end{figure}

\begin{table}[H]
\centering
\caption{Comparison of Variable Impact and Statistical Significance Across Groups and Scenarios. When comparing group $G_{i}$ with $G_{i+1}$, a "+" indicates a statistically significant higher value in $G_{i+1}$, while a "-" denotes a lower value. A green color signifies a favorable outcome for $G_{i+1}$ relative to $G_{i}$, and red indicates an adverse effect. The +/- sign indicates a difference in the distribution.}
\label{tab:significance}
\begin{tabular}{@{}lcccc@{}}
\toprule
Variable & Scenario & Group G1 vs G2 & Group G2 vs G3 & Group G3 vs G4   \\ \midrule
Accuracy & S1       &                &                &                               \\
         & S2       &                &      &       \textbf{\textcolor{green}{+}}   \\
         & S3       &  \textbf{\textcolor{red}{-}}   &                &                  \\
Alarms Acknowledged & S1 &             &    \textbf{\textcolor{green}{+}}            & \textbf{\textcolor{red}{-}} \\
                    & S2 &             &                &                               \\
                    & S3 &\textbf{\textcolor{green}{+}}      &   &                 \\
Alarms Silenced & S1    &              &                & \textbf{\textcolor{red}{-}}  \\
                & S2    &              &                &                               \\
                & S3    & \textbf{\textcolor{green}{+}}  &                &         \\
Mimics Opened  & S1-S3 &              &                &                               \\
Number of Alarms & S1  & \textbf{\textcolor{green}{-}}  &  & \textbf{\textcolor{green}{-}} \\
                 & S2  &              &        & \textbf{\textcolor{green}{-}}  \\
                 & S3  &              &                &                               \\
Reaction Time   & S1   &    &   & \textbf{\textcolor{green}{-}}\\
                & S2   &    &   & \textbf{\textcolor{green}{-}}\\
                & S3   &              &                &                               \\
Recovery Time   & S1   &              &                &  \textbf{\textcolor{green}{-}}   \\
                & S2   &              &                &  \textbf{\textcolor{green}{-}}  \\
Response Time   & S1   &              &\textbf{\textcolor{green}{-}}   & \\
                & S2   &              &                & \textbf{\textcolor{green}{-}}  \\
                & S3   & \textbf{\textcolor{red}{+}} &     &  \textbf{\textcolor{green}{-}}\\
Number of procedure  & S1   &              &  & \textbf{\textcolor{green}{-}} \\
                & S2   &              &                &   \\
                & S3   &  &     & \\              
Consequence   & S1   &              &    &                               \\
                & S2   &              &                &  \textbf{\textcolor{green}{+}}\\
                & S3   &               &                &               \\
Overall performance   & S1   &              &      &    \textbf{\textcolor{green}{+}}    \\
                & S2   &              &   +/-             & \textbf{\textcolor{green}{+}}          \\
Error rate   & S1   &             &        &          \\
            & S2    &             &        &          \\
            & S3    &    \textbf{\textcolor{green}{-}}          &        &          \\              
\bottomrule
\end{tabular}
\end{table}

\section{Prediction - Results and Discussion}

In this section, we develop and compare two predictive models, step-wise Logistic regression and Bayesian network, explicitly based on the behavioural metrics and the scenario and group variables from the operational data to forecast potential outcomes: Success or Failure. This enables the identification of key patterns and trends that may influence outcomes. Upon establishing the model, we leverage it to extract valuable insights, facilitating a deeper understanding of the operational dynamics and their impact on safety, efficiency, and overall performance. The extracted information proves instrumental in informing decision-making processes, enhancing operational strategies, and mitigating risks associated with the forecasted consequences.  The predictors considered included only the behavioural variables from previously discussed metrics such as the "Reaction Time", "Response Time", "Alarms Acknowledged", "Mimics Opened", "Number of Alarms", and the "Number of Procedures". These have been selected because of the possibility to extract them in a more practical control room setting. The "Scenario" (also seen as the Task Complexity) and "Group" (also seen as the type of support configuration) were also included.

\subsection{Prediction - Bayesian Network}

In this study, we extended our analysis to encompass both the NB model and the TAN model for the predictive assessment of operational outcomes. This dual approach enables us to leverage the intrinsic advantages of Bayesian network methodologies in understanding and modeling the complexities inherent in operational data. We use HUGIN EXPERT software to build and assess the Bayesian network. The operational variables considered in our research include reaction time, total procedure count, alarm silencing, alarm acknowledgment, mimic opening, accuracy, response time, total alarm count, and classifications by group and scenario. We use scenario as a complexity variable and the group as differents level of support. These variables are critical in constructing a comprehensive framework for our predictive analysis.

To validate the robustness and applicability of our models across varied operational settings, we employed a Stratified Shuffle Split cross-validation strategy with 1000 folds. Additionally, we utilized an entropy-based discretization technique for the effective categorization of continuous variables into discrete intervals. Given the small size of the dataset and the variability introduced by entropy discretization on the training set of each fold, the performance fluctuates. Therefore, a large number of folds were used to ensure the stability and reliability of the results.

The performance of the models was evaluated using a confusion matrix and several key metrics. The confusion matrix indicates the distribution of predictions as shown in \cref{tab:confusion_matrix_tan} and \cref{tab:confusion_matrix_nb}:

\begin{figure}[H]
\centering
\includegraphics[scale=0.8]{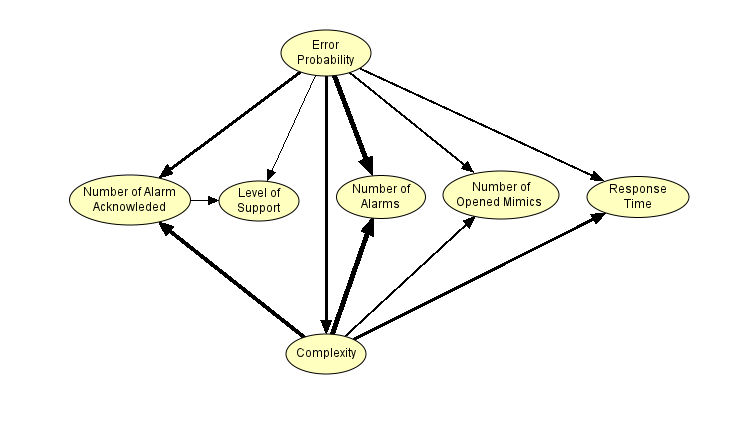}
\caption{TAN model structure to predict operator error. Most connections originate from the root variable "Complexity", with the exception of "Acknowledgements" being connected to "Level of Support". The size of the directed edges is proportional to the symmetric mutual information between the variables.} \label{TAN}
\end{figure}

\begin{table}[H]
\centering
\caption{Average on 1000 folds of the Confusion Matrix for TAN Model Performance}
\begin{tabular}{ccc}
\toprule
 & \textbf{Predicted Negative} & \textbf{Predicted Positive} \\
\midrule
\textbf{Actual Negative} & 55.54 & 3.46 \\
\textbf{Actual Positive} & 5.10 & 17.89 \\
\bottomrule
\end{tabular}
\label{tab:confusion_matrix_tan}
\end{table}


A threshold of 50\% was chosen for the classification of control room operator error. The model achieved a precision of 0.838, indicating that 83.8\% of the predicted positive instances were true positives. The accuracy of 0.895 shows that 89.5\% of all predictions were correct, while a recall of 0.778 demonstrates the model's effectiveness in identifying 77.8\% of actual positive instances. The F1 score of 0.807 balances precision and recall, confirming the model's solid performance in predicting human errors. Also, the AUC (Area Under the Curve) is 0.934. This level of performance suggests that the TAN model is effective in supporting decision-making processes in the control room by accurately predicting potential human errors based on metrics like "Recovery Time," "Reaction Time," "Response Time," "Accuracy," "Alarms Acknowledged," "Mimics Opened," and "Number of Alarms."

In comparison, the Naive Bayes model attained a precision of 0.781, accuracy of 0.876, recall of 0.773, an F1 score of 0.777 and AUC of 0.921. Indicating that the Naive Bayes model has solid but slightly lower performance compared to the TAN model in predicting human errors. The confusion matrix for the NB model is as follows:

\begin{table}[H]
\centering
\caption{Confusion Matrix for Naive Bayes Model Performance}
\begin{tabular}{ccc}
\toprule
 & \textbf{Predicted Negative} & \textbf{Predicted Positive} \\
\midrule
\textbf{Actual Negative} & 54.02 & 4.98 \\
\textbf{Actual Positive} & 5.21 & 17.79 \\
\bottomrule
\end{tabular}
\label{tab:confusion_matrix_nb}
\end{table}

Interestingly, the TAN model did not reveal many strong correlations between variables, indicating that while it captures more complex dependencies than the Naive Bayes model, the added complexity does not significantly enhance performance in this context. Most feature variables are directly connected to the root node "Scenario," highlighting its critical role in influencing the other variables, as shown in the model structure provided \cref{TAN}. In this figure we can also see the strength of the relation between the variables by its width. The widths of the link are proportional to the symmetric mutual information. 

Overall, both models demonstrate strong predictive capabilities. However, the slight performance edge of the TAN model underscores the importance of considering variable dependencies in probabilistic models. This comparison also highlights that the simpler Naive Bayes model remains a viable and efficient tool for similar predictive tasks, especially when ease of implementation and computational efficiency are prioritized.

\subsection{BN Model analysis}

The Tree Augmented Naive Bayes (TAN) model, having demonstrated superior performance in our comparative analysis, warrants a more in-depth examination. This section delves into the intricacies of the TAN model, leveraging mutual information analysis to uncover the relative importance of various factors in predicting operator errors. We'll also explore how the model estimates error probabilities across different scenarios and operator groups, providing crucial insights into the effectiveness of decision support systems in complex control room environments. Through this analysis, we aim to shed light on the key drivers of operator performance.

\subsubsection{Mutual information}

In our study, mutual information (MI) was utilized to assess the dependency between the operational variables and the target variable 'Error'. The MI values indicate the reduction in uncertainty about the 'Error' variable given the knowledge of the predictor variable. Higher MI values denote a stronger relationship between the predictor and the target variable.

\begin{figure}
\centering
\includegraphics[scale=0.8]{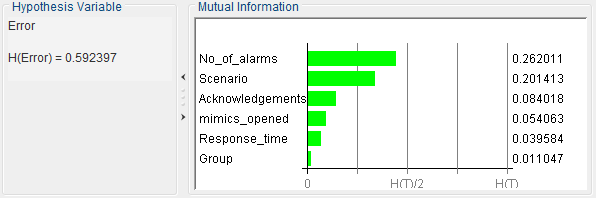}
\caption{Mutual information between the error and the feature variables. The entropy of the error H(error) can be seen on the right.} \label{MI}
\end{figure}

The analysis, as depicted in Figure \cref{MI}, highlights \textit{No. of alarms} with the highest MI score of 0.26. This indicates that the number of alarms is a critical factor influencing the error variable, implying a strong dependency and suggesting that a higher number of alarms is significantly associated with errors in operations.

\textit{Scenario} follows with an MI score of 0.20, suggesting that different operational scenarios contribute significantly to the prediction of errors. This emphasizes the need to consider scenario-specific factors when analyzing operational performance and predicting errors.

\textit{Acknowledgements} and \textit{mimics opened} also show notable MI values (0.08 and 0.05, respectively). These variables indicate that the rate of alarm acknowledgments and the number of mimic displays opened during operations have a meaningful impact on the error variable, though to a lesser extent compared to \textit{No. of alarms} and \textit{Scenario}.

On the other hand, \textit{Response time} and \textit{Group} present lower MI scores (0.04 and 0.01, respectively). The low score for \textit{Group} suggests a minimal direct impact on the error variable within the context of this study. Similarly, the relatively low MI for \textit{Response time} indicates that while response time affects the error variable, its influence is less significant compared to the number of alarms and scenario conditions.

It is important to note that \textit{Reaction time} is not present in the MI analysis. This absence is due to the application of the entropy-based discretization algorithm. The algorithm determines optimal cut points for continuous variables by maximizing the class-attribute interdependence. In this case, the algorithm could not find any threshold for Reaction time that would significantly improve the prediction of the error variable. This indicates that \textit{Reaction time} has no discernible connection with the error variable, reinforcing its lack of predictive power within the context of our study.

In summary, the mutual information analysis identifies \textit{No. of alarms} and \textit{Scenario} as the most significant predictors of the error variable. These insights are crucial for developing targeted strategies to minimize errors in operations. By focusing on these key variables, it is possible to enhance operational performance and reduce the incidence of errors.

\subsubsection{Outcome per group and scenario}
\cref{tab:probability_} presents the probability of success for each group across three different scenarios (S1, S2, and S3). The success probabilities provide insight into the effectiveness of different operational setups and technologies in achieving desired outcomes under varying conditions.

    Scenario 1 (S1): All groups demonstrated a high probability of success, with Group 1, Group 2, and Group 3 each achieving a 0.99 probability of success, and Group 4 slightly lower at 0.98. This indicates that under less challenging conditions, the presence of alarm prioritization, digital procedures, or AI decision support does not significantly affect the probability of success, as all groups perform at a near-optimal level.

    Scenario 2 (S2): The differences between groups become more apparent in this scenario. Group 2, which uses alarm prioritization, showed a substantial improvement with a probability of success at 0.87, compared to Group 1’s 0.69. Group 3, with digital procedures, had a success probability of 0.86, slightly lower than Group 2 but still significantly higher than Group 1. Group 4, which incorporates AI decision support, achieved the highest success probability at 0.91. These results suggest that alarm prioritization, digital procedures, and AI decision support all contribute to improved performance in moderately challenging conditions, with AI providing the most significant benefit.

    Scenario 3 (S3): This scenario presents the most challenging conditions, where the probability of success drops considerably across all groups. Group 1, without any advanced systems, has the lowest success probability at 0.22. Group 2’s success probability improves to 0.52, indicating that alarm prioritization offers a significant advantage. Group 3’s probability is 0.36, showing that the use of the digital procedure increased the error rate compared to paper-based. Group 4, with AI support, has a success probability of 0.31, which, while better than Group 1, is lower than both Groups 2 and 3. This suggests that under highly challenging conditions, alarm prioritization with paper-based procedures provides the greatest benefit, and the addition of AI support does not perform as well as expected, potentially due to the complexity of the scenario or the need for further refinement of AI tools.

Overall, the data indicates that while all advanced systems (alarm prioritization, digital procedures, and AI decision support) improve success probabilities, their effectiveness varies depending on the scenario’s complexity. The DSS group performed better in moderate complexity and the paper-based procedure with alarm rationalization performed better in complex scenarios.

\begin{table}[H]
\centering
\caption{Probability of Success for Each Group Across Scenarios for each group and scenario according to the TAN model built }
\label{tab:probability_}
\begin{tabular}{@{}lcccc@{}}
\toprule
\textbf{Scenario} & \textbf{Group 1} & \textbf{Group 2} & \textbf{Group 3} & \textbf{Group 4} \\
\midrule
S1 & 0.99 & 0.99 & 0.99 & 0.98 \\
S2 & 0.69 & 0.87 & 0.86 & 0.91 \\
S3 & 0.22 & 0.52 & 0.36 & 0.31 \\
\bottomrule
\end{tabular}

\end{table}

\subsection{Prediction - Logistic Regression (LR)}

\subsubsection{Step 1: }
The first step in the subsection involved predicting the outcome [Error: Success, Failure] based on all variables, including the type of group and scenario. Predictions were made in Python language using Jupyter Notebook.


\begin{table*}[ht]
    \centering
    \caption{Coefficients and statistical significance of predictors using Logistic Regression. All groups and scenarios combined and without feature selection. \textit{Accuracy: 86\%, Precision: 76.9\%, Recall: 71\%, F1 Score: 74\%, ROC-AUC Score: 92\% }}
    \begin{tabular}{
        >{\raggedright\arraybackslash}p{5cm} 
        S[table-format=1.4] 
        S[table-format=1.3] 
        S[table-format=-1.3] 
        S[table-format=1.3] 
        S[table-format=-1.3, table-space-text-pre={[}, table-space-text-post={]}] 
    }
    \toprule
    \textbf{Predictor} & {\textbf{Coefficient}} & {\textbf{Std Err}} & {\textbf{z-value}} & {\textbf{\(P>|z|\)}}   \\
    \midrule
    Group            & -2.1600 & 0.551 & -3.923 & 0.000  \\
    Scenario         & 3.8938 & 1.182 & 3.295 & 0.001\\
    Reaction\_time   & -1.3770 & 0.877 & -1.570 & 0.116  \\
    Response\_time   & -2.5546 & 1.079 & -2.368 & 0.018  \\
    Acknowledgements & -1.2167 & 1.123 & -1.083 & 0.279  \\
    Mimics\_opened    & -2.7103 & 1.028 & -2.637 & 0.008  \\
    No\_of\_alarms    & 3.4684 & 1.294 & 2.681 & 0.007  \\
    \bottomrule
    \end{tabular}
    \label{tab:coefficients}
\end{table*}

Compared to the outcome in \cref{MI}, using the logistic regression technique, the best variables with respect to highest coefficient impact are Scenario, No of alarms, acknowledgement, reaction time, group, response time, and mimics opened (see \cref{tab:coefficients}). 




The confusion matrix and derived metrics indicate that the model performs quite well overall with an Accuracy of 86\%, Precision: 76.9\%, Recall: 71\%, F1 Score: 74\%, and ROC-AUC Score: 92\% . Overall, the high values of precision, recall, F1 and AUC score demonstrate that the model is effective at distinguishing between the classes, especially given the high recall, which indicates strong performance in identifying actual positive cases. The slightly lower precision compared to recall suggests there are some false positives, but this is balanced by the high recall rate.

However, the relevant variables only should be considered for prediction as this has the advantage of providing a better model. To achieve this, a step-wise logistic regression is applied.

\subsubsection{Performing a step-wise logistic regression}
After performing a step-wise logistic regression with a forward-feature selection, the key variables according to highest coefficient impact are 'Scenario', 'No. of alarms', 'Group', Mimics opened' and 'Response time'. Using this approach produced a slightly better  ROC-AUC score of 93\% compared to the previous. 

Also, this time around, acknowledgement were removed as it was not considered important which is not the case in \cref{MI} when BN was used. Similar to \cref{MI}, Reaction time is not considered significant to predict the error in the LR algorithm. The mutual information in \cref{MI} considered the group of less impact, which is not the case in the step-wise logistic outcome. However, in both cases, No alarms, scenarios, mimics opened and response time, were significant.

%
. 
    
Both outcomes in step 1 and 2 indicate good performance, but using the step-wise logistic regressions can streamline the amount of variables included and potentially improve performance.

\subsubsection{Outcome per group and scenario with LR}
 Given that the type of scenario/event and group impacts significantly the outcomes and behaviour of the variables, it would be important to see how each group in each scenario perform as similarly shown in \cref{tab:probability_}. To calculate the probability of failure for each group and scenario, same step wise logistic model was employed as before, using the same selected features: number of alarms, response time,mimics opened, Group and Scenario. For each data point, the probability of failure was calculated using the logistic function:
\begin{equation}
p(failure) = \frac{1}{1 + e^{-(\beta_0 + \beta_1x_1 + \beta_2x_2 + \beta_3x_3 + \beta_4x_4)}}
\end{equation}
where $\beta_i$ are the model coefficients and $x_i$ are the standardized feature values.
Then the probabilities are aggregated by group and scenario, calculating the mean probability for each combination. Table \ref{tab:my_label} presents the resulting probabilities of failure for each group across the three scenarios. These results indicate that Scenario S3 consistently presents the highest risk of failure across all groups, with probabilities ranging from 0.55 to 0.69. Group\_1 shows the highest overall probability of failure, particularly in Scenario S3 (0.69), while Group\_4 demonstrates the lowest overall probability of failure across all scenarios. Also, Group\_3 and Group\_4 had close outcomes in terms of failure probability across the three scenarios.

\begin{table}[ht]
    \centering
    \caption{Probability of success for each group in scenario 2 and 3 according to the step-wise logistic regression model. Higher percentage of Success predictions compared to the TAN model (see fig \ref{MI}).}
    \label{tab:my_label}
    \begin{tabular}{ccccc}
        \Xhline{2pt}
        Scenario & Group 1 & Group 2 & Group 3 & Group 4 \\
        \hline
        S1 & 0.94 & 0.96 & 0.96 & 0.97 \\
        \hline
        S2 & 0.79 & 0.84 & 0.87 & 0.89 \\
        \hline
        S3 & 0.31 & 0.38 & 0.43 & 0.45 \\
        \Xhline{2pt}
    \end{tabular}
    
\end{table}





\subsection{Predicting with subjective measures}
As an extra step towards exploring the importance of classical performance shaping factors on alarm response outcomes, some of the subjective variables measured during the experiment, such as training, familiarity, situational awareness and workload, were included. The situational awareness techniques used for the questionnaire-based data collection are the situational awareness rating technique (SART) and the situation presence assessment method (SPAM), while that of workload is the NASA Task Load Index (NASA-TLX). When included in the dataset and run for the combined group and scenario comparison using the step-wise logistic regression technique, the result in Table \ref{tab:additional_coefficients} was realised. Some behavioural variables were not significant for prediction following the inclusion of these subjective variables. Also, training and familiarity were not significant for the prediction. 

\begin{table*}[ht]
    \centering
    \caption{Coefficients and statistical significance of additional predictors. \textit{Accuracy: 84\%, Precision: 71.00\%, Recall: 71\%, F1 Score: 71\% and AUC: 91.7\%, at threshold of 0.5}
}
    \label{tab:additional_coefficients}
    \begin{tabular}{
        >{\raggedright\arraybackslash}p{5cm} 
        S[table-format=1.4] 
        S[table-format=1.3] 
        S[table-format=-1.3] 
        S[table-format=1.3] 
        S[table-format=-1.3, table-space-text-pre={[}, table-space-text-post={]}] 
    }
    \toprule
    \textbf{Predictor} & {\textbf{Coefficient}} & {\textbf{Std Err}} & {\textbf{z-value}} & {\textbf{\(P>|z|\)}} \\
    \midrule
    Scenario         & 3.8900 & 0.964 & 4.036 & 0.000 \\
    Response\_time   & -2.7665 & 0.998 & -2.771 & 0.006  \\
    TLX\_Index       & 3.4879 & 1.313 & 2.657 & 0.008 \\
    SART\_Index      & -7.6300 & 1.151 & -6.627 & 0.000 \\
    \bottomrule
    \end{tabular}

\end{table*}

Incorporating the new subjective variables, as detailed in Table \ref{tab:additional_coefficients}, identified SART Index, Response time, Scenario, and TLX Index as key predictors. Variables such as the number of alarms, reaction time, group, mimics opened, acknowledgment, SPAM Index, familiarity, and training were not chosen by the step wise LR model. Nevertheless, it remains evident that Scenario and response time consistently proved to be significant indicators of error in both earlier and recent predictions. Moreover, the importance of situational awareness and workload for correct alarm response was confirmed by the significance of SART and NASA TLX for the error prediction as shown in Table \ref{tab:additional_coefficients}. These findings highlight the necessity of considering both behavioral and cognitive measures when evaluating operator performance.

\section{Comparison of Bayesian Network and Logistic Regression Approaches}

This study employed both BN and LR techniques to predict operator errors in control room environments. While both methods demonstrated strong predictive capabilities, they offered unique insights and advantages. This section compares their performance and discusses the strengths of each approach.

Both the TAN model and the step-wise LR model showed high predictive accuracy. Table \ref{tab:performance_comparison} presents a comparison of their performance metrics.

\begin{table}[ht]
\centering
\caption{Performance Comparison of BN and LR Models}
\label{tab:performance_comparison}
\begin{tabular}{lcc}
\hline
Metric & TAN Model & Step-wise LR Model \\
\hline
Accuracy & 89.5\% & 86\% \\
Precision & 83.8\% & 76.9\% \\
Recall & 77.8\% & 71\% \\
F1 Score & 80.7\% & 74\% \\
ROC-AUC & 93.4   & 93\% \\
\hline
\end{tabular}
\end{table}

\cref{tab:performance_comparison} presents a comparative analysis of the performance metrics between the TAN model and the Step-wise LR model. Both models demonstrate strong predictive capabilities, with the TAN model slightly outperforming the LR model across all metrics. The TAN model achieves higher accuracy (89.5\% vs. 86\%), precision (83.8\% vs. 76.9\%), recall (77.8\% vs. 71\%), and F1 score (80.7\% vs. 74\%). The ROC-AUC scores are nearly identical, with the TAN model at 93.4 and the LR model at 93\%. These results indicate that while both models are effective in predicting operator errors, the Bayesian Network approach (TAN) provides marginally better overall performance, particularly in terms of precision and recall. This suggests that the TAN model may be slightly more reliable in identifying control room operator error and minimizing false positives.

The two methods identified similar key predictors but with some notable differences. Table \ref{tab:variable_importance} summarizes the most important variables identified by each method.

\begin{table}[ht]
\centering
\caption{Ranked Key Predictors Identified by BN and LR Models}
\label{tab:variable_importance}
\begin{tabular}{ll}
\hline
TAN Mutual Information Analysis & Step-wise LR \\
\hline
1. No. of alarms & 1. Scenario \\
2. Scenario & 2. No. of alarms \\
3. Acknowledgements & 3. Group \\
4. Mimics opened & 4. Response time \\
5. Response time &  5. Mimics opened \\
6. Group &  \\
\hline
\end{tabular}
\end{table}

Both of the models place the Number of alarms and scenario/task complexity as the main factors influencing control room operator error. The LR model placed more importance on the 'Group/type of support' variable, which was less significant in the BN analysis. Conversely, the BN model emphasized 'Acknowledgements', which was not selected in the final LR model. Also, both models did not include reaction time as an important predictor.

\subsection{Advantages of Each Method}
The Bayesian Network approach offers several distinct advantages. It captures complex dependencies, allowing the TAN model to represent more nuanced relationships between variables. BNs can handle missing data, making predictions even with incomplete information. They provide probabilistic output, offering more nuanced insights through probability distributions for outcomes. The graphical structure of BNs allows for an intuitive understanding of variable relationships, enhancing interpretability.

Logistic Regression also presents unique strengths. It offers simplicity, being easier to implement and interpret, especially for non-technical stakeholders. LR provides direct impact measurement, with coefficients directly indicating the impact of each variable on the outcome. The step-wise LR method enables efficient variable selection, automatically selecting the most relevant predictors. Additionally, LR is robust to outliers, being less sensitive to extreme values compared to some other methods.

\subsection{Discussion}

The high performance of both models suggests that either could be effectively deployed in a control room setting. The choice between them might depend on specific operational needs. If interpretability and ease of implementation are priorities, the LR model might be preferable. If capturing complex variable interactions and handling missing data are crucial, the BN approach could be more suitable.

The slight differences in identified important variables between the two methods highlight the value of using multiple analytical approaches. For instance, the LR model's emphasis on the 'Group' variable suggests that the type of support system might have a more direct impact on error rates than initially apparent from the BN analysis.

In conclusion, both methods offer valuable insights into predicting operator errors. The BN approach provides a more comprehensive probabilistic framework, while the LR method offers a straightforward, easily interpretable model. Using both in tandem can provide a more robust understanding of the factors influencing operator performance in control room environments.

\section{Conclusion}

In this paper, we have demonstrated through the operational data the best configurations, with focus on a combination of certain situational awareness and decision support tools, when dealing with varying complexities of safety-critical tasks.


Alarm prioritization demonstrated clear benefits, particularly in scenarios with high alarm intensity. It led to improved alarm management, higher acknowledgment rates, and better overall performance. The impact was most pronounced in complex situations, suggesting that alarm prioritization is a valuable tool for managing cognitive load and focusing operator attention on critical issues.

The transition from paper-based to digital procedures showed mixed results. In less complex scenarios, digital procedures improved response times and potentially reduced alarm clutter, indicating enhanced efficiency. They also provided more consistent performance across operators, reducing the variability seen with paper-based procedures. However, the impact was less significant in highly complex scenarios, and in some cases, led to less efficient alarm management. This suggests that while digital procedures offer benefits, their implementation should be carefully considered and may require additional operator training to fully leverage their potential.

The integration of AI-based DSS showed promising results in low to moderate complexity scenarios. It improved decision accuracy, reduced reaction and response times, and facilitated faster problem resolution. The DSS also appeared to streamline operations by reducing the need for extensive procedure consultation. However, its effectiveness was limited in highly complex scenarios, highlighting the need for further development to address challenging situations.

The analysis of operational data across different scenarios and support configurations revealed intriguing patterns in operator performance and error rates. Mutual information analysis identified the number of alarms and scenario complexity as the most significant predictors of errors, emphasizing the critical role of alarm management and scenario-specific factors in operational safety. The effectiveness of various support tools varied notably across different complexity levels. While all advanced systems generally improved success probabilities, their impact was not uniform across scenarios. Notably, in the most challenging conditions (Scenario 3), alarm prioritization with paper-based procedures (Group 2) showed the highest probability of success, outperforming both digital procedures and AI-based decision support systems. This unexpected finding suggests that in highly complex situations, the familiarity and simplicity of paper-based systems, combined with effective alarm prioritization, may offer advantages over more technologically advanced solutions. These results underscore the importance of tailoring support systems to specific operational contexts and complexity levels, rather than assuming that more advanced technology will always yield better outcomes.

In conclusion, the study demonstrates the value of exploring operational data to understand operators' behaviour and its impact on safety by comparing two machine learning algorithms: step-wise logistic regression (LR) and Bayesian Networks (BN). Their potential for error prediction in safety-critical contexts is identified. Behavioural metrics, such as reaction time, response time, number of alarms acknowledged, and number of mimics opened, alongside established metrics like the number of alarms and situational awareness, workload, play a crucial role. It is important to note that the list of possible metrics considered here is not exhaustive, as there remain unaccounted percentages needed to achieve more robust predictions. This leaves room for further research to explore a wider range of possible variables, including those derived from biometric and neurometric tools used during data collection.

A limitation of this study is the use of static models. Future research will focus on employing dynamic models, such as the incremental step-wise logistic regression algorithm, to enhance predictive accuracy and adaptability.

\section*{Consent and Ethics Statement}

Participants were briefed prior to the study, read a detailed description of what the study entailed via an information sheet, and signed the necessary consent form before participating. These documents, together with the ethics application, were approved by the Internal Ethical Committee of the Collaborative Intelligence for Safety-Critical Systems, following a first approval by the Ethics Review Committee of the Technological University of Dublin Ireland with approval number REC-20-52.

\section*{Funder Information and Acknowledgments}

This work has been carried out within the Collaborative Intelligence for Critical Safety Systems (CISC) project. The CISC project has received funding from the European Union's Horizon 2020 Research and Innovation Programme under Marie Skodowska-Curie grant agreement no. 955901. The authors thank Rob Turner and Adrian Kelly in Yokogawa and EPRI Europe for contributing to the development of the human-system interfaces. Finally, thanks to the participants for the time they dedicated to this study and their feedback.

\section*{Declaration of Interest}

The authors declare that they have no known competing financial interests or personal relationships that could have appeared to influence the work reported in this paper.

\section{Data Availability Statement}

Preliminary data supporting this study's findings are available on GitHub at \url{https://github.com/CISC-LIVE-LAB-3/dataset}.

\section{Declaration of generative AI and AI-assisted technologies in the writing process}
During the preparation of this work the author(s) used Claude AI from Anthropic to clarify long sentences and correct spelling. After using this tool/service, the author(s) reviewed and edited the content as needed and take(s) full responsibility for the content of the published article.


\bibliographystyle{elsarticle-num}
\bibliography{Manuscript}

\begin{thebibliography}{10}
\expandafter\ifx\csname url\endcsname\relax
  \def\url#1{\texttt{#1}}\fi
\expandafter\ifx\csname urlprefix\endcsname\relax\def\urlprefix{URL }\fi
\expandafter\ifx\csname href\endcsname\relax
  \def\href#1#2{#2} \def\path#1{#1}\fi

\bibitem{Simonson2022}
R.~J. Simonson, J.~R. Keebler, E.~L. Blickensderfer, R.~Besuijen, \href{https://doi.org/10.1016/j.apergo.2021.103670}{{Impact of alarm management and automation on abnormal operations: A human-in-the-loop simulation study}}, Applied Ergonomics 100~(May 2021) (2022) 103670.
\newblock \href {https://doi.org/10.1016/j.apergo.2021.103670} {\path{doi:10.1016/j.apergo.2021.103670}}.
\newline\urlprefix\url{https://doi.org/10.1016/j.apergo.2021.103670}

\bibitem{Gao2013}
Q.~Gao, Y.~Wang, F.~Song, Z.~Li, X.~Dong, {Mental workload measurement for emergency operating procedures in digital nuclear power plants}, Ergonomics 56~(7) (2013) 1070--1085.
\newblock \href {https://doi.org/10.1080/00140139.2013.790483} {\path{doi:10.1080/00140139.2013.790483}}.

\bibitem{Hinss2022}
M.~F. Hinss, A.~M. Brock, R.~N. Roy, {Cognitive effects of prolonged continuous human-machine interaction: The case for mental state-based adaptive interfaces}, Frontiers in Neuroergonomics 3 (2022).
\newblock \href {https://doi.org/10.3389/fnrgo.2022.935092} {\path{doi:10.3389/fnrgo.2022.935092}}.

\bibitem{Gertman2004}
D.~I. Gertman, H.~S. Blackman, J.~L. Marble, C.~Smith, R.~L. Boring, P.~O'Reilly, {The SPAR H human reliability analysis method}, American Nuclear Society 4th International Topical Meeting on Nuclear Plant Instrumentation, Control and Human Machine Interface Technology (2004) 17--24.

\bibitem{Iqbal2024}
M.~U. Iqbal, B.~Srinivasan, R.~Srinivasan, {Multi-class classification of control room operators' cognitive workload using the fusion of eye-tracking and electroencephalography}, Computers and Chemical Engineering 181~(December 2023) (2024).
\newblock \href {https://doi.org/10.1016/j.compchemeng.2023.108526} {\path{doi:10.1016/j.compchemeng.2023.108526}}.

\bibitem{Braarud2021}
P.~O. Braarud, T.~Bodal, J.~E. Hulsund, M.~N. Louka, C.~Nihlwing, E.~Nystad, H.~Svengren, E.~Wingstedt, \href{https://doi.org/10.1177/0018720820961730}{An investigation of speech features, plant system alarms, and operator–system interaction for the classification of operator cognitive workload during dynamic work}, Human Factors 63~(5) (2021) 736--756, pMID: 33054415.
\newblock \href {http://arxiv.org/abs/https://doi.org/10.1177/0018720820961730} {\path{arXiv:https://doi.org/10.1177/0018720820961730}}, \href {https://doi.org/10.1177/0018720820961730} {\path{doi:10.1177/0018720820961730}}.
\newline\urlprefix\url{https://doi.org/10.1177/0018720820961730}

\bibitem{Rahman2020}
H.~Rahman, M.~U. Ahmed, S.~Barua, S.~Begum, \href{https://doi.org/10.1016/j.bspc.2019.101634}{{Non-contact-based driver's cognitive load classification using physiological and vehicular parameters}}, Biomedical Signal Processing and Control 55 (2020) 101634.
\newblock \href {https://doi.org/10.1016/j.bspc.2019.101634} {\path{doi:10.1016/j.bspc.2019.101634}}.
\newline\urlprefix\url{https://doi.org/10.1016/j.bspc.2019.101634}

\bibitem{Chen2012}
F.~Chen, N.~Ruiz, E.~Choi, J.~Epps, M.~A. Khawaja, R.~Taib, B.~Yin, Y.~Wang, {Multimodal behavior and interaction as indicators of cognitive load}, ACM Transactions on Interactive Intelligent Systems 2~(4) (2012).
\newblock \href {https://doi.org/10.1145/2395123.2395127} {\path{doi:10.1145/2395123.2395127}}.

\bibitem{9557368}
B.~Schmidt, R.~Borrison, M.~Gärtler, S.~Maczey, A.~Kotriwala, Practical aspects for exploration and analysis of manual interventions in process plants, in: 2021 IEEE 19th International Conference on Industrial Informatics (INDIN), 2021, pp. 1--6.
\newblock \href {https://doi.org/10.1109/INDIN45523.2021.9557368} {\path{doi:10.1109/INDIN45523.2021.9557368}}.

\bibitem{CROMPTON202183}
J.~Crompton, \href{https://www.sciencedirect.com/science/article/pii/B9780128207147000054}{Chapter 5 - data management from the dcs to the historian}, in: P.~Bangert (Ed.), Machine Learning and Data Science in the Oil and Gas Industry, Gulf Professional Publishing, 2021, pp. 83--110.
\newblock \href {https://doi.org/https://doi.org/10.1016/B978-0-12-820714-7.00005-4} {\path{doi:https://doi.org/10.1016/B978-0-12-820714-7.00005-4}}.
\newline\urlprefix\url{https://www.sciencedirect.com/science/article/pii/B9780128207147000054}

\bibitem{Demichela2017}
M.~Demichela, G.~Baldissone, G.~Camuncoli, Risk-based decision making for the management of change in process plants: benefits of integrating probabilistic and phenomenological analysis, Industrial \& Engineering Chemistry Research 56~(50) (2017) 14873--14887.

\bibitem{AmazuDOE2024}
C.~W. Amazu, J.~Mietkiewicz, A.~N. Abbas, H.~Briwa, A.~Alonso-Perez, G.~Baldissone, D.~Fissore, M.~Demichela, M.~C. Leva, \href{https://doi.org/10.1080/10447318.2024.2376354}{Exploring the influence of human system interfaces: Introducing support tools and an experimental study}, International Journal of Human–Computer Interaction 0~(0) (2024) 1--18.
\newblock \href {http://arxiv.org/abs/https://doi.org/10.1080/10447318.2024.2376354} {\path{arXiv:https://doi.org/10.1080/10447318.2024.2376354}}, \href {https://doi.org/10.1080/10447318.2024.2376354} {\path{doi:10.1080/10447318.2024.2376354}}.
\newline\urlprefix\url{https://doi.org/10.1080/10447318.2024.2376354}

\bibitem{mietkiewicz2024enhancing}
J.~Mietkiewicz, A.~N. Abbas, C.~W. Amazu, G.~Baldissone, A.~L. Madsen, M.~Demichela, M.~C. Leva, Enhancing control room operator decision making, Processes 12~(2) (2024) 328.

\bibitem{AMAZUDatainBrief2024}
C.~W. Amazu, J.~Mietkiewicz, A.~N. Abbas, H.~Briwa, A.~{Alonso Perez}, G.~Baldissone, M.~Demichela, D.~Fissore, A.~L. Madsen, M.~C. Leva, \href{https://www.sciencedirect.com/science/article/pii/S2352340924001410}{Experiment data: Human-in-the-loop decision support in process control rooms}, Data in Brief 53 (2024) 110170.
\newblock \href {https://doi.org/https://doi.org/10.1016/j.dib.2024.110170} {\path{doi:https://doi.org/10.1016/j.dib.2024.110170}}.
\newline\urlprefix\url{https://www.sciencedirect.com/science/article/pii/S2352340924001410}

\bibitem{Shi2022}
C.~Shi, L.~Rothrock, \href{https://doi.org/10.1080/00140139.2022.2132299}{{Using eye movements to evaluate the effectiveness of the situation awareness rating technique scale in measuring situation awareness for smart manufacturing}}, Ergonomics 0~(0) (2022) 1--9.
\newblock \href {https://doi.org/10.1080/00140139.2022.2132299} {\path{doi:10.1080/00140139.2022.2132299}}.
\newline\urlprefix\url{https://doi.org/10.1080/00140139.2022.2132299}

\bibitem{Yang2021}
J.~Yang, N.~Liang, K.~O. Prakah-Asante, R.~Curry, M.~Blommer, R.~Swaminathan, B.~J. Pitts, D.~Yu, {Situation Awareness Classification Using Multi-modal Sensing in Automated Driving}, Proceedings of the Human Factors and Ergonomics Society 65~(1) (2021) 52.
\newblock \href {https://doi.org/10.1177/1071181321651095} {\path{doi:10.1177/1071181321651095}}.

\bibitem{Solovey2014}
E.~T. Solovey, M.~Zec, E.~A.~G. Perez, B.~Reimer, B.~Mehler, {Classifying driver workload using physiological and driving performance data: Two field studies}, Conference on Human Factors in Computing Systems - Proceedings (2014) 4057--4066\href {https://doi.org/10.1145/2556288.2557068} {\path{doi:10.1145/2556288.2557068}}.

\bibitem{jensen2007bayesian}
F.~V. Jensen, T.~D. Nielsen, Bayesian Networks and Decision Graphs: February 8, 2007, Springer, 2007.

\bibitem{Anders2013}
U.~Kjærulff, A.~Madsen, Bayesian Networks and Influence Diagrams: A Guide to Construction and Analysis, Springer, 2013.

\bibitem{fayyad1993multi}
U.~M. Fayyad, K.~B. Irani, Multi-interval discretization of continuous-valued attributes for classification learning, in: Ijcai, Vol.~93, Citeseer, 1993, pp. 1022--1029.

\bibitem{liu2002discretization}
H.~Liu, F.~Hussain, C.~L. Tan, M.~Dash, Discretization: An enabling technique, Data mining and knowledge discovery 6 (2002) 393--423.

\bibitem{shapiro1965analysis}
S.~S. Shapiro, M.~B. Wilk, An analysis of variance test for normality (complete samples), Biometrika 52~(3/4) (1965) 591--611.

\bibitem{levene1960robust}
H.~Levene, Robust tests for equality of variances, in: Contributions to Probability and Statistics: Essays in Honor of Harold Hotelling, Stanford University Press, 1960, pp. 278--292.

\bibitem{student1908probable}
Student, The probable error of a mean, Biometrika 6~(1) (1908) 1--25.

\bibitem{welch1947generalization}
B.~L. Welch, The generalization of ‘student's’ problem when several different population variances are involved, Biometrika 34~(1-2) (1947) 28--35.

\bibitem{mann1947test}
H.~B. Mann, D.~R. Whitney, On a test of whether one of two random variables is stochastically larger than the other, The Annals of Mathematical Statistics 18~(1) (1947) 50--60.

\end{thebibliography}

\section*{About the Authors}

\textbf{Chidera Winifred Amazu}, a Ph.D. student and research assistant at Politecnico di Torino, is part of the Collaborative Intelligence for Safety-Critical Systems network. Her research centres on the HMI for situational awareness, decision support of control room operators, and safety analysis. She studies operator behaviour and cognition during human-machine interaction.

\textbf{Joseph Mietkiewicz}, with a Master’s in Mathematics and ongoing PhD studies at Hugin Expert Denmark and TUDublin Ireland, contributes to the Collaboration Intelligence for Critical Safety Systems project. Specializing in AI, particularly interpretable machine learning and Bayesian networks, he focuses on developing decision support systems for control room operators.

\textbf{Ammar N. Abbas}, with a master’s in Mechatronics and pursuing a PhD in Deep Reinforcement Learning, focuses on optimal decision-making in safety-critical systems (detecting anomalies and prescriptive maintenance, optimising the product quality, process scheduling.) He utilizes Reinforcement Learning as an online method, integrating human expertise with explainable and interpretable AI.

\textbf{Gabriele Baldissone}, is currently a Research Technician and Lecturer at Politecnico di Torino, Italy. He educates undergraduate and Master’s students in advanced control and environmental safety techniques, as well as advanced technologies for risk-based decision-making. His expertise spans risk analysis, assessment, modelling, quantitative analysis, reliability, data mining, and process safety. 

\textbf{Davide Fissore}, Full Professor of Process Control and Food Processing Technologies at Politecnico di Torino, specialises in process modelling and development of advanced model-based tools for process monitoring and control. His recent research extends to pharmaceutical engineering, focusing on applying the 'Quality by Design' concept to optimise, monitor and control pharmaceutical freeze-drying processes.

\textbf{Micaela Demichela}, Full Professor at Politecnico di Torino, is actively involved in teaching and research and holds positions on the Board of Directors for the ESReDa Association and the Scientific Board for the R3C Inter-Department Centre. Her expertise lies in risk management, occupational safety, probabilistic risk assessment, etc.

\textbf{Anders L. Madsen}
Anders L Madsen, Full Professor at Department of Computer Science at Aalborg University (part time) and Chief Executive Officer at HUGIN EXPERT A/S contributes to the Collaboration Intelligence for Critical Safety Systems project. His expertise is in algorithms for Bayesian networks and applications of Bayesian network.

\textbf{Maria Chiara Leva}, co-chair of the Human Factors Technical Committee for the European Safety and Reliability Association, previously chaired the Irish Ergonomics Society and co-chaired the Symposium on Human Mental Workload. She co-founded Tosca Solutions and currently lectures at TU Dublin. Her expertise spans Human Factors and Safety Management Systems.

\end{document}